\documentclass[aps,prd,twocolumn,superscriptaddress,nofootinbib,showpacs,letterpaper]{revtex4}
\usepackage{graphicx}
\usepackage{amsmath}
\usepackage{amsfonts}
\usepackage{mathrsfs}
\usepackage{ulem}
\usepackage[latin1]{inputenc}
\usepackage[usenames,dvipsnames]{color}
\newcommand{\Caltech}{\affiliation{Theoretical Astrophysics 350-17, California Institute of Technology, Pasadena, CA 91125}}
\newcommand{\WVU}{\affiliation{Department of Physics, West Virginia University, PO Box 6315, Morgantown, WV 26506}}
\newcommand{\beq}{\begin{equation}}
\newcommand{\eeq}{\end{equation}}
\newcommand{\bea}{\begin{eqnarray}}
\newcommand{\eea}{\end{eqnarray}}
\newcommand{\ba}{\begin{align}}
\newcommand{\ea}{\end{align}}

\newcommand{\bma}{\begin{pmatrix}}
\newcommand{\ema}{\end{pmatrix}}

\begin{document}
\title{A joint approach for reducing eccentricity and 
spurious gravitational radiation
\\ in binary black hole initial data construction} 

\author{Fan Zhang} \Caltech \WVU
\author{B\'ela Szil\'agyi} \Caltech

\begin{abstract}
  At the beginning of binary black hole simulations, there
  is a pulse of spurious radiation (or junk radiation) resulting from the initial data not
  matching astrophysical quasi-equilibrium inspiral exactly.  One traditionally
  waits for the junk radiation to exit the computational domain before taking physical
  readings, at the expense of throwing away a segment of the evolution, and
  with the hope that junk radiation exits cleanly.  We argue that this hope does not
  necessarily pan out as junk radiation could excite long-lived constraint violation.
  Another complication with the initial data is that it contains orbital
  eccentricity that needs to be removed, usually by evolving the early
  part of the inspiral multiple times with gradually improved input parameters.
  We show that this procedure is also adversely impacted by junk radiation. In this paper,
  we do not attempt to eliminate junk radiation directly, but instead tackle the much
  simpler problem of ameliorating its long-lasting effects. We report on the
  success of a method that achieves this goal by combining the removal of junk radiation
  and eccentricity into a single procedure. Namely we periodically stop a low resolution
  simulation; take the numerically evolved
  metric data and overlay it with eccentricity
  adjustments; run it through initial data solver (i.e. the solver receives as free data  
  the numerical output of the previous iteration); 
  restart the simulation;
  repeat until eccentricity becomes sufficiently low, and then launch the high
  resolution ``production run'' simulation. This approach has the following
  benefits: (1) We do not have to contend with the influence of junk radiation on eccentricity 
  measurements for
  later iterations of the eccentricity reduction procedure. (2) We re-enforce constraints
  every time initial data solver is invoked, removing the constraint violation
  excited by junk radiation
  previously. (3) The wasted simulation segment associated with the junk radiation's 
  evolution is absorbed into the eccentricity reduction iterations. 
  Furthermore, (1) and (2) together allow us to carry
  out our joint-elimination procedure at low resolution, even when the
  subsequent ``production run'' is intended as a high resolution simulation.
\end{abstract}
\date{\today}
\maketitle

\section{Introduction \label{sec:Intro}}

Binary black hole coalescences constitute one of the most 
promising sources for the gravitational wave detectors
such as Advanced LIGO \cite{Harry2010}, Virgo \cite{aVIRGO,aVIRGO:2012} and
KAGRA \cite{Somiya:2011np}. In order to achieve a detection, 
we need numerical simulations to calibrate and validate the 
template bank of waveforms. 

When we prepare the initial data to be used in such a binary black hole simulation,
we could not obtain an exact snapshot of a
quasi-equilibrium inspiral. 
Once the simulation begins, the spacetime relaxes into a 
 quasi-equilibrium configuration with the
mismatch radiating away as a pulse of spurious or ``junk" radiation
(JR).
JR, therefore, can be thought of as the gravitational
  perturbation
  one needs to add to the quasi-equilibrium solution in order to obtain
the solution used as numerical initial data.  Part of this perturbation
will propagate outwards immediately as outgoing gravitational radiation.
Another part corresponds to an ingoing radiation that results in a small 
but long lasting transient \cite{Bishop:2011iu}.
Yet another part will fall onto the black holes, exciting quasi-normal
ringing which, in turn, will result in outgoing
gravitational waves of much higher frequency than those generated
by the orbital motion of the binary
\cite{Lovelace2009,Boyle2007}. 
 
Resolving the highest frequency components of JR, although
possible, is not practical as our numerical grid is tuned for resolving the
quasi-equilibrium system, where all short ($\approx1M$) length-scale features
are in the immediate neighborhood of the binary, while the grid in the outer
regions is built to resolve wave propagation on much larger ($\approx100M$)
length-scales.  The alternative approach had been to just accept the fact JR is
not resolved and wait for it to leave the computational
domain \cite{Lovelace2009}.  The result is that the JR may morph into
constraint violation (CV) that can alter the physical properties of the
system (individual masses and spins, orbital parameters), and degrade 
the accuracy of the simulation for
longer than one light-crossing time of the computational domain.

Another complication with constructing initial data is the astrophysically
motivated need for low orbital eccentricity (OE).  Given that gravitational
radiation tends to circularize binary orbits
\cite{PetersMathews1963,Peters1964,Postnov:2007jv}, it is expected that
for binaries born from stellar evolution, rather than e.g. dynamical capture
scenarios \cite{East:2012xq}, little OE would remain by the time the system enters
into the sensitive band of the next generation of ground-based gravitational
wave detectors. 
It is therefore desirable to remove OE in simulations meant to generate
waveforms for the template banks of these detectors. This entails evolving the
simulation for a small number of orbits and reading off the oscillation in the
orbital angular velocity of the black holes. Then this data is used to generate an
improved set of initial data parameters. Such
a procedure is repeated a number of times until OE reduces to an acceptable
level. Junk radiation complicates this procedure by introducing a long-lasting
disturbance into the eccentricity measurement of the binary. 

In this paper, we aim to reduce such long-lasting secondary effects created by JR. 
To this end, we
simply modify the traditional approach where one waits for the JR to exit the
outer boundary before taking physical readings \footnote{ 
Instances of taking such physical readings include extracting
  gravitational wave signals, comparing with post-Newtonian waveforms
\cite{Boyle2007,Hannam2007}, or calculating kick velocity in a binary black
hole system with recoil \cite{Choi-Kelly-Boggs-etal:2007,Gonzalez2007}.}. 
In our approach, at the end of each eccentricity reduction iteration,
we make use of the numerically evolved metric data, apply adjustments on the
black holes' velocities, and then re-solve the constraints before continuing on,
 rather than adjusting
the initial data parameters and starting the run from the same point as
in the initial iteration.

After the initial iteration of this ``stop-and-go'' operation, the JR will have
passed through the computational domain. Furthermore, by invoking the initial
data solver, we not only blend in the velocity adjustments, but also remove CV
built up previously. In other words, we resolve the issue that JR does not exit
cleanly. For later iterations, the eccentricity measurement of the binary is no longer
perturbed by the JR, allowing for a more robust eccentricity reduction procedure.
After a few iterations, we would have converged onto a snapshot of an
astrophysically realistic inspiral that has reduced JR, OE, as well as CV, which
can be used as the initial data \footnote{This new cleaner initial data would
have evolved into different physical parameters from the one we had at the very
beginning, with the differences reasonably predicted by post-Newtonian formulae
(e.g. see \cite{Campanelli2007b} for spin precession frequency), because we are at
the early part of an inspiral.}
for a high resolution ``production run'' simulation. From this perspective, we
have jointly eliminated OE and the impact of JR with a numerical process.
We thus refer to this collection of ``stop-and-go''
iterations as the ``joint-elimination'' approach. 

It is worth mentioning that we regard our method as being complementary to the
  analytical approach, where one targets the JR itself by obtaining a more
  realistic initial data through adding more analytical components into it
  \cite{JohnsonMcDaniel:2009dq}, such as wave content \cite{kellyEtAl:2007}, 
  tidal deformations, or allowing for conformally curved initial data
  \cite{JohnsonMcDaniel:2009dq,Lovelace2009}.  These algorithms are aimed at
  reducing the JR in what we refer to as the initial iteration 
  (to be followed by first, second etc iterations) of the 
  joint-elimination approach.  The resulting
  quasi-equilibrium system has a better chance of closely matching the
  desired physical parameters.  However, inherent to these analytic
  approaches, one must employ some sort of blending, truncated expansion, etc.
  The resulting initial data will be different from the quasi-equilibrium system,
  and will generate some amount of JR.  
  Our approach then takes whatever outcome is available from the analytical procedures, 
evolve it
through the JR phase, and construct a subsequent evolution that is now based on
the numerically achieved quasi-equilibrium solution.

Our study is carried out with the Spectral Einstein Code (\verb!SpEC!)
\cite{SXSWebsite}, a pseudospectral code that solves a first differential order
form of the generalized harmonic formulation of Einstein's equations
\cite{Lindblom:2007,Friedrich1985,Garfinkle2002,Pretorius2005c}. The overall domain of evolution
is divided into the so-called ``subdomains'' that are simply spherical shells
near the excision boundaries and in the gravitational wave zone, and are
more complicated shapes in the inner
regions \cite{Scheel2009,Szilagyi:2009qz}.  

We begin by summarizing some useful background information in
Sec.~\ref{Sec:Background}, before moving on to elaborate on the junk radiation's
long-lasting impacts in Sec.~\ref{Sec:problems}. In Sec.~\ref{sec:Method}, we
describe the joint-elimination procedure, and then in Sec.~\ref{Sec:Result}, 
we demonstrate its effectiveness and advantages. 

In the formulae below, the early part of
the Latin alphabet denotes spacetime indices that run from 0 to 3, while the
mid part of the Latin alphabet denotes spatial indices (from 1 to 3). Bold face
letters represent tensors or vectors. Unless stated otherwise, the figures
depict the inspiral stage of an equal-mass nonspinning black hole binary
simulation (referred to as the ``standard example''), beginning from conformally
flat initial data with an initial coordinate separation of $15$M (M being the
total mass). 
This simulation is carried out on the overlapping
subdomain decomposition of \cite{Scheel2009,Szilagyi:2009qz}, at the lowest resolution
typically used in a convergence study. 

\section{Preliminaries \label{Sec:Background}}
In this section, we review some of the formalisms used in initial data
construction, eccentricity reduction and spacetime curvature visualization. Our
joint-elimination procedure will use a variant of the methods summarized here. 
The adaptations will be summarized in
Sec.~\ref{sec:Method}. 

\subsection{Vacuum initial value problem\label{Sec:IDSum}}
In the standard $3+1$ form of Einstein's equations \cite{ADM,york79}, the metric
is written as 
\bea
ds^2 = -N^2 dt^2 + g_{ij}\left( dx^i +  N^i dt \right)\left( dx^j +  N^j dt \right)
\eea
where $N$ is the lapse, $N^i$ is the shift, and $g_{ij}$ is the three metric
on a constant $t$ hypersurface. 

The extrinsic curvature of that hypersurface is given by 
\begin{equation} \label{eq:DefExCur}
{\bf K} = - \frac{1}{2} \mathbb{P}\mathcal{L}_{{\bf n}} {}^{(4)}{\bf g}
\end{equation}
where ${}^{(4)}\bf{g}$ is the spacetime metric,  $\mathcal{L}_{{\bf n}}$ is the Lie
derivative along the unit normal to the hypersurface, and $\mathbb{P}$ is the
projection operator into the hypersurface. The definition
Eq.~\eqref{eq:DefExCur} leads to
\begin{equation} \label{eq:ExCurvVsShift}
\partial_t g_{ij} = -2N K_{ij} + \nabla_i N_j + \nabla_j N_i,
\end{equation}
so $K_{ij}$ gives the time derivative of $g_{ij}$ aside from a shift correction,
and constitutes the central piece in the Arnowitt-Deser-Misner canonical
momentum \cite{ADM}
\bea
\pi^{ij} = \sqrt{g}\left( K^{ij} - K g^{ij}\right).
\eea

Specifying $g_{ij}$ and $K_{ij}$ then amounts to pinning down the initial state
in phase space, or in other words providing the initial data. In addition, one
should also specify the gauge choice by fixing the initial values for lapse $N$
and shift $N^i$.   

The initial data is not arbitrary. It has to satisfy the 
Hamiltonian and momentum constraints 
\begin{eqnarray}
R+K^2-K_{ij}K^{ij}&=&0, \label{eq:HamCons} \\
\nabla_j\left(K^{ij}-g^{ij}K\right)&=&0, \label{eq:MomCons}
\end{eqnarray}
where $\nabla_j$ is the covariant derivative on the spatial hypersurface, and
$R$ is the trace of the spatial Ricci tensor. 
These constraints represent the condition that $K_{ij}$ and $g_{ij}$ are
consistent with being associated with a slice of a \textit{vacuum} spacetime. 
In a geometrical sense these constraints
enforce that the spatial surface with prescribed $K_{ij}$ and $g_{ij}$ can
indeed be immersed into a four dimensional Ricci-flat ambient spacetime. 

Constraint satisfying initial data construction is done by designating a set of
functions as freely specifiable while leaving the rest determined by the constraint 
equations. We have one scalar and one vector equation in
Eqs.~\eqref{eq:HamCons} and
\eqref{eq:MomCons} respectively, so we should leave one scalar plus one vector
quantity indeterminate. We assign them to intrinsic metric $g_{ij}$ and
extrinsic curvature $K_{ij}$ respectively, and choose the quantities that the
constraint equations depend on sensitively. Following the approach of the extended
conformal thin sandwich formalism (XCTS) \cite{York1999,Pfeiffer2003b}, we define 
\bea
g_{ij} = \psi^4 \tilde{g}_{ij} .
\eea
We specify the conformal spatial metric $\tilde{g}_{ij}$ by hand and solve for
the conformal factor $\psi$ using essentially the Hamiltonian constraint
Eq.~\eqref{eq:HamCons}.  We will denote with a tilde quantities associated with
the conformal metric.  For the momentum constraint Eq.~\eqref{eq:MomCons}, we
first decompose extrinsic curvature to its trace (i.e. mean curvature) and the
transverse traceless and longitudinal traceless parts
\begin{equation} \label{eq:ShiftToExCurv}
K^{ij} = \frac{1}{3}g^{ij} K + M^{ij} + \frac{1}{2N}(\mathbb{L}V)^{ij}
\left(\equiv \frac{1}{3}g^{ij} K + A^{ij}\right)
\end{equation}
that have the following conformal counterparts 
\bea
\tilde{K} &=& K, \quad \tilde{M}^{ij} = \psi^{10} M^{ij}, \notag \\
\quad (\tilde{\mathbb{L}}\tilde{V})^{ij} &=& \psi^4 (\mathbb{L}V)^{ij}, \quad
\tilde{A}_{ij} = \psi^2 A_{ij}.
\eea
The longitudinal operator is defined as 
\bea
(\mathbb{L}V)^{ij} = 2 \nabla^{(i}V^{j)} - \frac{2}{3} g^{ij} \nabla_k V^k, 
\eea
and $V^i$ can serve as the indeterminate vector field. Comparing
Eq.~\eqref{eq:ShiftToExCurv} with Eq.~\eqref{eq:ExCurvVsShift}, we are not
surprised to find that when the shift gauge condition is set to $N^i = V^i$, the
conformal $\tilde{M}_{ij}$ simply becomes the (lapse weighted) trace-free part
of the time derivative of the conformal metric, which we denote as
$\tilde{u}_{ij}$. 
Under this particular gauge fixing, momentum constraint takes on the pretense of
an equation for shift (with some $\psi$ coupled in), even though it is really solving for extrinsic curvature.

We can now rewrite the Hamiltonian and momentum constraints as a pair of coupled
elliptic equations for $\psi$ and $N^i$.  
With XCTS, one also sets the lapse gauge condition by specifying $\partial_t
\tilde{K}$, which translates into an equation for the conformal lapse $\tilde{N}$
\cite{Pfeiffer2003b,Smarr78b,Frontiers:Wilson}. The combined system of equations is then 
\begin{widetext}
\bea \label{eq:CTS}
\tilde{\nabla}^2 \psi - \frac{1}{8} \psi \tilde{R} - \frac{1}{12} K^2 \psi^5 +
\frac{1}{8} \tilde{A}_{ij} \tilde{A}^{ij} \psi^{-7} &=& 0, \\
\tilde{\nabla}_j \left( \frac{1}{2\tilde{N}} (\tilde{\mathbb{L}}\tilde{N})^{ij}
\right) - \frac{2}{3} \psi^6 \tilde{\nabla}^i K - \tilde{\nabla}_j
\left(\frac{\tilde{u}^{ij}}{2\tilde{N}} \right) &=& 0, \\
\tilde{\nabla}^2 \left(\tilde{N} \psi^7\right) - \left(\tilde{N}
\psi^7\right)\left[ \frac{\tilde{R}}{8} + \frac{5}{12} K^4 \psi^4 + \frac{7}{8}
\psi^{-8} \tilde{A}^{ij} \tilde{A}_{ij}\right] &=& -\psi^5 \left(\partial_t K -
N^k \partial_k K\right),
\eea
\end{widetext}
where $\tilde{N} = \psi^{-6} N$. We solve these equations with a
multidomain spectral elliptic partial differential equation solver described in
\cite{Pfeiffer2003}. The pre-determined free-data are the conformal metric
$\tilde{g}_{ij}$, the trace-removed part of its time derivative
$\tilde{u}_{ij}$, the mean curvature $K$ and its time derivative $\partial_tK$.
Generating realistic initial data means providing better
values for these quantities. 

In order to provide sensible free-data, it is convenient to go into a
co-rotating frame where the black holes are pinned down at fixed spatial
coordinates. There, one can impose the quasi-stationary conditions
$\tilde{u}_{ij} = 0$ and $\partial_t K = 0$. The remaining free-data that do
not involve time derivatives can be either $\tilde{g}_{ij} = \eta_{ij}$ 
($\bf{\eta}$ being the Minkowski spatial metric) and
$K=0$ in the conformally flat case, or, for the superposed Kerr-Schild case
\cite{Lovelace2009} (more conducive to high spins), a combination of
Kerr-Schild  black holes inside Gaussian envelopes
\begin{eqnarray} \label{eq:SKSFreeData}
\tilde{g}_{ij} &=& \eta_{ij} + \sum_{a=1}^2 e^{-r_a^2/\omega_a^2} (g_{ij}^a -
  \eta_{ij})  \\
K &=& \sum_{a=1}^2 e^{-r_a^2/\omega_a^2} K_a 
\end{eqnarray}
where $g^a$ and $K_a$ are spatial metric and mean curvature of spinning
Kerr-Schild black holes. 

The boundary conditions on the unknown variables are also important, as for
example in the conformally flat case, the presence and properties of the black
holes are completely fixed by the boundary conditions. The shift condition on
the excision boundaries in the co-rotating frame is 
\bea
N^i = N s^i + N^i_{\parallel}
\eea
where $s^i$ is normal to the boundaries and $N^i_{\parallel}$ tangential to it,
giving the spin of the black holes. There is no component directly corresponding
to the orbital motion of the black holes. That piece of information enters
through the boundary condition at spatial infinity $r \rightarrow \infty$ ($r$
is the magnitude of the location vector ${\bf r}$) 
\begin{equation} \label{eq:RotationalShiftRotOnly}
N^i = \xi^i \equiv ({\bf \Omega}_0 \times {\bf r})^i 
\end{equation}
After solving for $(\psi_{\text{co}}, N_{\text{co}}, N_{\text{co}}^i)$ in the
co-rotating frame, translating into the inertial frame solution
$(\psi_{\text{in}}, N_{\text{in}}, N_{\text{in}}^i)$ is straightforward. It
turns out that since $\mathbb{L} {\bf \xi} = 0$ \cite{Pfeiffer-Brown-etal:2007}, we have
that, provided mean curvature $K$ vanishes, the quantities 
\bea
\psi_{\text{in}}=\psi_{\text{co}}, \quad
N^i_{\text{in}} = N^i_{\text{co}}, \quad
N_{\text{in}}^i = N^i_{\text{co}} - \xi^i, 
\eea
would satisfy the XCTS equations with the inertial frame boundary conditions of 
\begin{equation} \label{eq:BCInertialFrameRotOnly}
N^i = N s^i - \xi^i + N^i_{\parallel}
\end{equation}
on the excision boundaries, as well as $N^i = 0$ at $r \rightarrow \infty$. For
the boundary conditions on $\psi$ and $N$, and a more thorough introduction to
XCTS formalism, see e.g. \cite{Pfeiffer-Brown-etal:2007}.  

\subsection{A recursive eccentricity reduction procedure \label{Sec:EcRed}}
As discussed in Sec.~\ref{sec:Intro}, binary black hole initial data construction usually
include an eccentricity reduction stage. The method we summarize here is a
recursive procedure proposed and developed in
\cite{Pfeiffer-Brown-etal:2007,Boyle2007,Tichy:2010qa,Buonanno:2010yk}, whereby 
initial data are evolved for two to three orbits, before an analysis of the orbit
in terms of the separation $s$ between the black hole apparent horizons, or
instantaneous angular velocity $\omega$, is carried out to generate an
improved set of initial parameters to be used in the next iteration. 

\begin{figure}[!t]
\includegraphics[width=0.85\columnwidth]{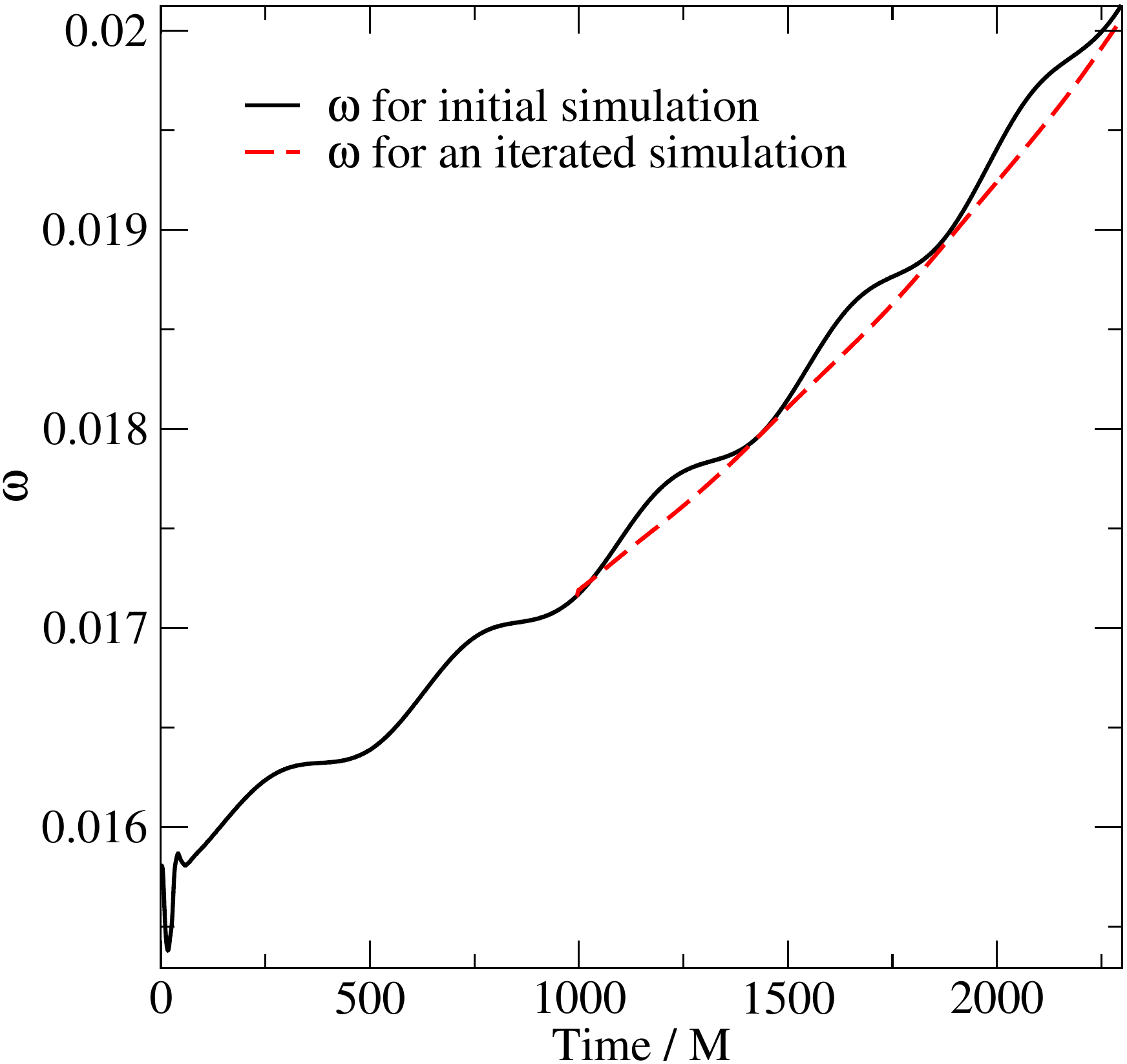}
\caption{The solid black curve represents the evolution of $\omega$ in the
standard example simulation. The wobbles signify the presence of eccentricity.
For comparison, we also display as a dashed red curve the $\omega(t)$ for this
simulation after one stop-and-go operation described in Sec.~\ref{sec:Method},
which contains an eccentricity suppression step. }
\label{fig:OmegaMag}
\end{figure}

Orbital eccentricity manifests itself as an oscillation in $ds/dt$ or
$d\omega/dt$ with a period close to the orbital period (but does not need to be
exactly equal when periastron advances are present \cite{Buonanno:2010yk}). 
To separate this oscillation from the smooth decline of orbital separation (the
``inspiral''), which is equivalently represented by an increase of orbital
frequency $\omega$ (see Fig.~\ref{fig:OmegaMag}), 
the time derivative $d\omega/dt$ (see Ref.~\cite{Buonanno:2010yk} for a discussion
on the advantage of using $d\omega/dt$ over $ds/dt$) is fitted to a functional form
\cite{Pfeiffer-Brown-etal:2007,Buonanno:2010yk}
\begin{eqnarray} \label{eq:Pfeiffer-Brown-etal:2007Fitting}
\frac{d\omega}{dt} &=& A_0(T-t)^{-11/8}+A_1(T-t)^{-13/8} \notag \\
&&+B \cos(\omega t + \phi + \nu t^2)
\end{eqnarray}
by varying parameters $(T,A_0,A_1,B,\phi,\nu)$. The first two terms constitute the
orbital decay, and the last one is the oscillation due to eccentricity. 

From the fitting result, one can calculate the eccentricity estimate 
\begin{equation} \label{eq:Pfeiffer-Brown-etal:2007Estimate}
e = \frac{B}{2\Omega_0 \omega}, 
\end{equation}
where $\Omega_0$ is the initial angular frequency. We can then calculate how
much tangential and radial velocities need to be added onto the black holes in order
to drive $e$ to zero. An initial radial velocity can be given to the black holes  
by changing the boundary condition
Eq.~\eqref{eq:BCInertialFrameRotOnly} on the excision surfaces in the inertial
frame to  
\begin{equation} \label{eq:BCInertialFrame}
N^i = N s^i - \xi_{\rm rad}^i + N^i_{\parallel},
\end{equation}
where 
\begin{equation} \label{eq:RotationalShift}
\xi_{\rm rad}^i \equiv ({\bf \Omega}_0 \times {\bf r})^i + \dot{a}_0 r^i. 
\end{equation}
The formulae (Eqs.~(74) and (78) of \cite{Buonanno:2010yk}) 
\begin{equation} \label{eq:Pfeiffer-Brown-etal:2007Deltas}
\delta \Omega_0 = - \frac{B}{4\Omega_0} \sin \phi, \quad \delta \dot{a}_0 = \frac{B}{2\Omega_0} \cos\phi
\end{equation}
then provide the adjustments to the orbital angular frequency $\Omega_0$ and
expansion factor $\dot{a}_0$ in Eq.~\eqref{eq:RotationalShift}, to be used 
for initial data construction in the next iteration. 
Note this is not a
simple change of gauge into one where the orbit looks superficially circular,
but instead a genuine change to physics because the extrinsic curvature is also
updated according to Eq.~\eqref{eq:ShiftToExCurv}. 

\subsection{Visualizing curvature \label{Sec:VisualSum}}
In order to study the anatomy of junk radiation, 
it is beneficial to be able to see the curvature structure
within the bulk of spatial slices. 
To this end, we construct visualization tools based on gauge invariant
contractions \cite{Carminati:1991} 
\begin{equation}
  I = \Psi_4 \Psi_0 - 4 \Psi_1 \Psi_3 + 3 \Psi_2^2, \quad J =
\left| \begin{array}{ccc}
\Psi_4 & \Psi_3 & \Psi_2 \\
\Psi_3 & \Psi_2 & \Psi_1 \\
\Psi_2 & \Psi_1 & \Psi_0 \end{array} \right|,
\end{equation}
of the Weyl tensor $C_{abcd}$, where $\Psi's$ are the Newman-Penrose scalars 
\bea
\Psi_0&=&-C_{abcd} l^a m^b l^c m^d, \\
\Psi_1&=&-C_{abcd} l^a n^b l^c m^d, \\
\Psi_2&=&-C_{abcd} l^a m^b \overline{m}^c n^d, \\
\Psi_3&=&-C_{abcd} l^a n^b \overline{m}^c n^d, \\
\Psi_4&=&-C_{abcd} n^a \overline{m}^b n^c \overline{m}^d.
\eea
extracted on any Newman-Penrose null tetrad $\{l_a,n_a,m_a,\overline{m}_a\}$
that consists of a pair of null vectors $({\bf l,n})$ and a complex conjugate
pair of complex vectors $({\bf m, \overline{m}})$, satisfying normalization
conditions such that the spacetime metric takes the standard form of  
\bea 
\bma
0 & -1 & 0 & 0 \\
-1 & 0 & 0 & 0 \\
0 & 0 & 0 & 1 \\
0 & 0 & 1 & 0 
\ema
\eea
on that tetrad basis. 
The expression 
\begin{equation} 
\Psi^t_2 = -\frac{1}{2}\left(P + \frac{I}{3P}\right)
\end{equation}
where 
\bea
P = \left[J + \sqrt{J^2-(I/3)^3}\right]^{1/3}, 
\eea
then gives the Coulomb background part of the Weyl tensor \cite{Szekeres1965,Baker:2000zm}.
Defining a pair of geometrical coordinates by \cite{QuasiKinn}
\begin{equation} 
\Psi^t_2 = M(\rho^t)^3, \quad 
\rho^t = \frac{1}{r^t - ia\cos(\theta^t)},
\end{equation}
we obtain a gauge invariant depiction of the structure of the Coulomb background
in the form of $(r^t,\theta^t)$ contours. For example, in the Kerr limit
$(r^t,\theta^t)$ become the Boyer-Lindquist coordinates whose contours, when
plotted in Kerr-Schild slicing and spatial coordinates, are asymptotically simple
spherical shells and cones (see Fig.~1 in \cite{QuasiKinn}). 

\section{Some features of junk radiation \label{Sec:problems}}
\subsection{A quadrupolar component of junk radiation \label{sec:RotatingQuadrupole}}
\begin{figure}[!t]
\includegraphics[width=0.85\columnwidth]{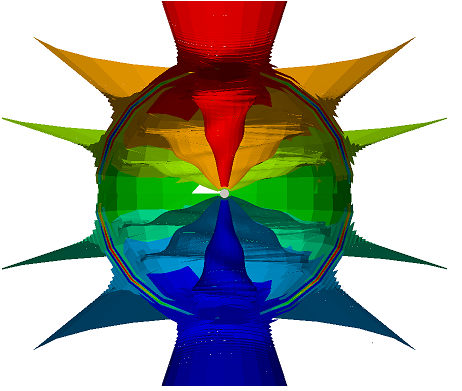}
\caption{Reproduced from \cite{QuasiKinn}. Visualization of junk radiation as the
boundary between unrealistic and realistic multipole structures in $\theta^t$
contours. }
\label{fig:JRVisual}
\end{figure}

The ability to see the curvature structure is useful for visualizing JR because
it marks the boundary between the regions containing unrealistic initial data and
the post-JR realistic data. As can be seen in the $\theta^t$ contours of 
Fig.~\ref{fig:JRVisual} (shown originally in \cite{QuasiKinn}), 
the region behind JR shows the signature spiral staircase pattern (see Fig.~2 in
\cite{QuasiKinn}) generated by a rotating mass quadrupole. Essentially the quadrupolar
moment squashes the cones of $\theta^t$ contour, and the rotation in this moment
causes the squashing direction to vary depending on the distance to the source region,
thus forming a twisting pattern. On the other hand, the region ahead of JR does
not contain the influence of the rotating quadrupolar moment, because the
superposed Kerr-Schild initial data being plotted does not correctly account for
inspiral history. Such an absence of past history is more typically seen in
Newtonian instantaneous action theories, and is a valid approximation in our
general relativistic context only when we are close to the black holes, where
the time retardation effect is insignificant. 

\begin{figure}[!t]
\includegraphics[width=0.95\columnwidth]{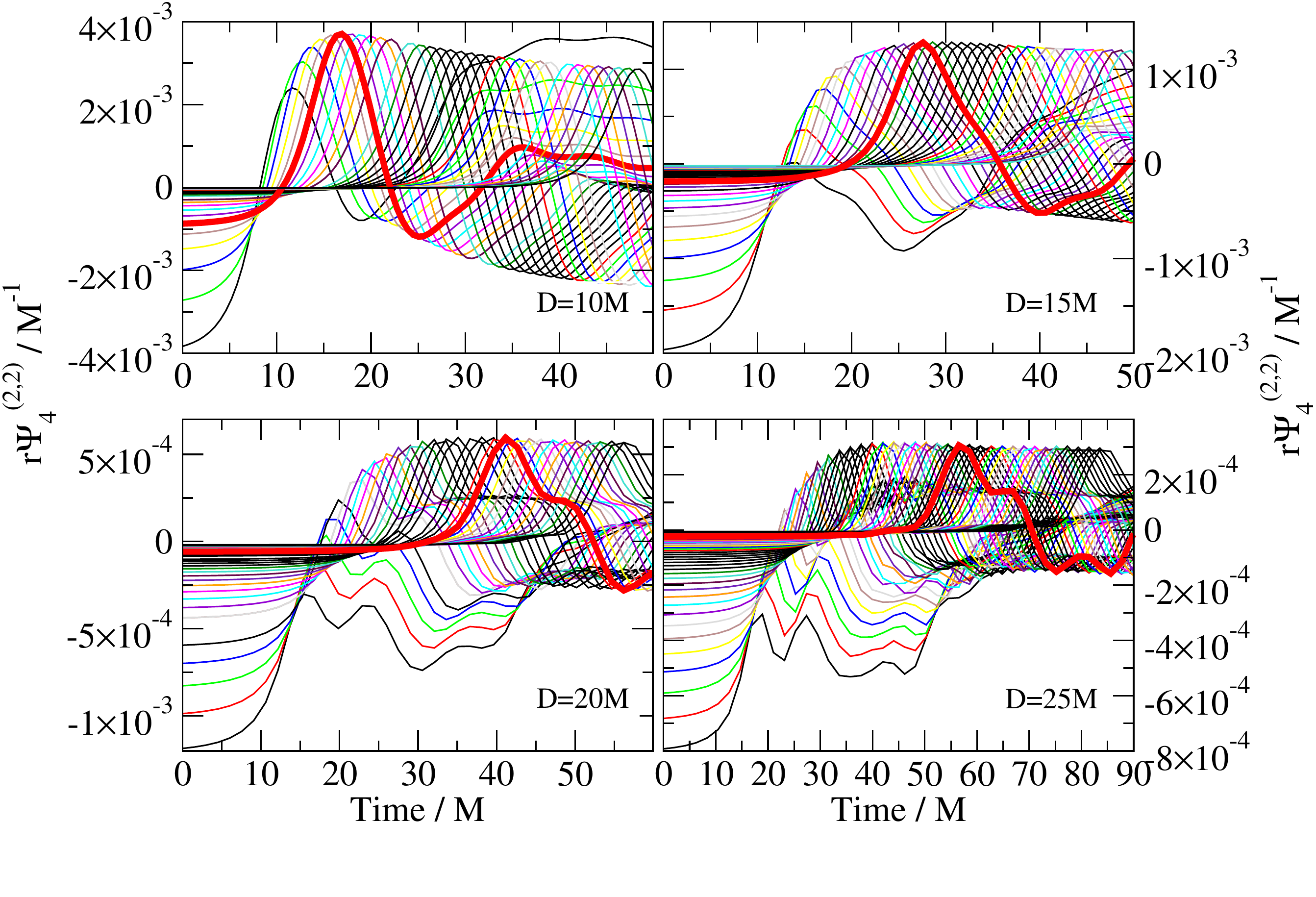}
\caption{The $(l=2,m=2)$ mode of $r\Psi_4$ waveform extracted at various
coordinate radii $r_{\text{ex}} \geq D + 1$M, with the thick red curve
representing $r_{\text{ex}} \approx \lambdabar$. Curves with peaks further to
the right correspond to larger $r_{\text{ex}}$. The four panels describe
simulations that are similar to the standard example but with initial
separations of $D=10$M, $15$M, $20$M and $25$M, and corresponding
$\lambdabar=15.8$M, $29.0$M, $44.7$M and $62.5$M. }
\label{fig:JRArrivalTime}
\end{figure}

To see whether this missing inspiral history represents the dominant omission in
initial data, we turn to the spatial distribution of JR magnitude. We begin by
noting that an approximate limit beyond which the instantaneous initial data
becomes invalid is given by 
\bea
r\agt\lambdabar= \frac{1}{2}\sqrt{\frac{D^3}{M}}
\eea
in Ref.~\cite{Nichols:2011pu} (see Figures 18 and 19 in that paper), where
$\lambdabar$ is the reduced gravitational wavelength with $D$ being the initial
binary separation and $M$ the total mass. The contribution from missing inspiral
history should begin picking up magnitude after JR reaches $r \approx
\lambdabar$, but those from near-zone dynamics should begin tapering off (i.e.
$r\Psi_4$ flattens) at this demarcation line between the near and transition
zones \cite{Nichols:2011pu}. We then carry out equal-mass nonspinning simulations
with initial separations of $D=10$M, $15$M, $20$M and $25$M, and corresponding
$\lambdabar=15.8$M, $29.0$M, $44.7$M and $62.5$M. In
Fig.~\ref{fig:JRArrivalTime}, we plot the $(l=2,m=2)$ mode of $r\Psi_4$ waveform
in these simulations extracted at various coordinate radii $r_{\text{ex}} \geq D
+ 1$M, with the thick red curve corresponding to $r_{\text{ex}} \approx
\lambdabar$. We concentrate on the early part of the waveforms and examine
whether magnitude of the junk radiation rose significantly for those extraction
radii satisfying $r_{\text{ex}} > \lambdabar$ (these are the curves with peaks
to the right of the thick red curve). The absence of this growth in JR magnitude
in Fig.~\ref{fig:JRArrivalTime} then suggests that the missing inspiral history
is unlikely the most significant source of JR. 

\subsection{Excitation of constraint violation \label{Sec:constraints}}
\begin{figure}[!b]
\includegraphics[width=0.9\columnwidth]{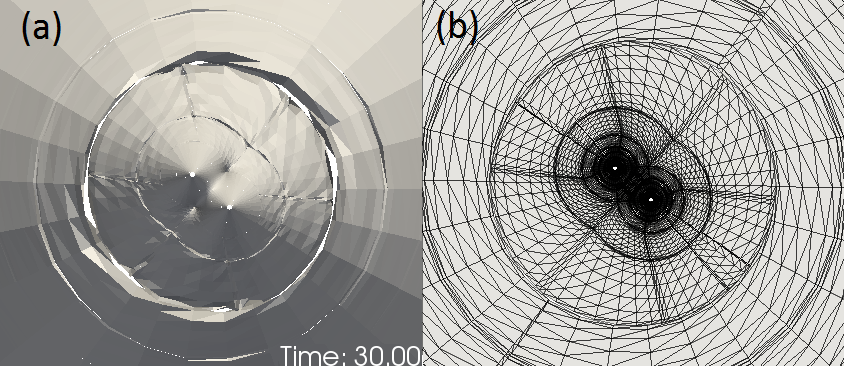}
\includegraphics[width=0.75\columnwidth]{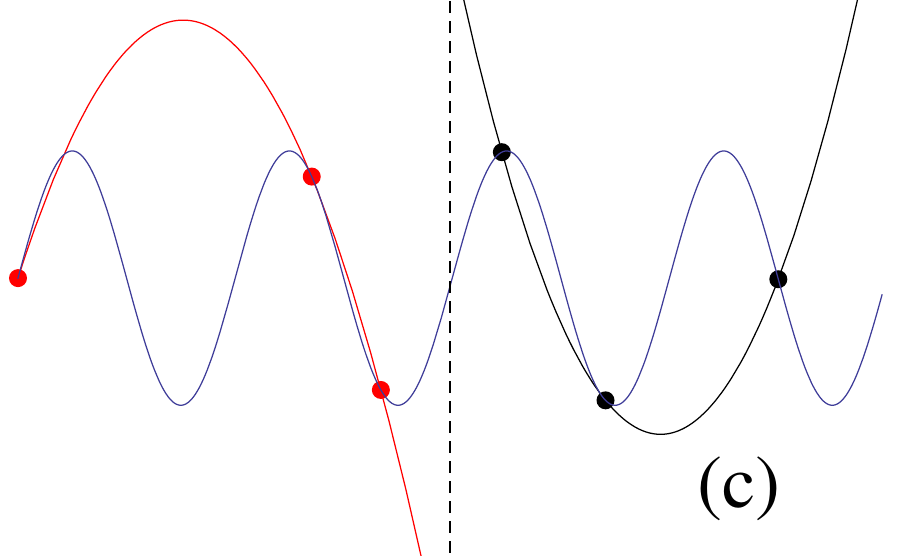}
\caption{(a) Equatorial plane slice of the computational domain in the standard
  example simulation (but with a non-overlapping subdomain decomposition for
  better visual clarity), warped into the paper according to $r^t$ value. The
  impact of JR on the subdomain boundaries is to tear them open in the fashion
  depicted in Panel (c). (b) The subdomain boundaries in (a) are shown as dense
  concentration of lines in this panel. Their location matches the tearing.
(c) A stylized demonstration that under-resolved high frequency JR (denoted by
blue sinusoidal curve) causes mismatch of data at subdomain boundary (vertical
dashed line). Red and black curves are the representations of the JR in the left
and right subdomains respectively, constructed from sampling at the
under-populated red and black dots.}
\label{fig:JRAtSubdomainBdry}
\end{figure}

By visualizing $r^t$, one can also learn interesting features of JR with regard
to generating constraint violation. The high frequency JR requires finer grids 
to resolve, which would cause the time step size to
also drop according to Courant limit. This is a very high computational cost, 
so even for simulations equipped with Adaptive Mesh Refinement \cite{Lovelace:2010ne}
(our simple standard example isn't), 
the refinement is left off in the gravitational wave zone 
until the JR has left the computational domain. The result is that
JR is under-resolved in all or some of the regions, and turns into constraint
violating modes that degrade the accuracy of the output.

One place where the creation of constraint violation is particularly visible is
at subdomain boundaries. The penalty method
\cite{Hesthaven1997,Hesthaven1999,Hesthaven2000,Gottlieb2001} adopted by some
pseudospectral codes such as \verb!SpEC! does not force the values of data
across the subdomain boundaries to match up exactly, so the high frequency
under-resolved JR will tear the boundary open, as is graphically explained in
Fig.~\ref{fig:JRAtSubdomainBdry} (c) and demonstrated for an actual simulation
in Fig.~\ref{fig:JRAtSubdomainBdry} (a), creating discontinuities and thus
constraint violation. Eventually the gap at the boundaries will be closed by
the penalty, but constraint violation will generally take longer to damp out. 

\begin{figure}[!t]
\includegraphics[width=0.99\columnwidth]{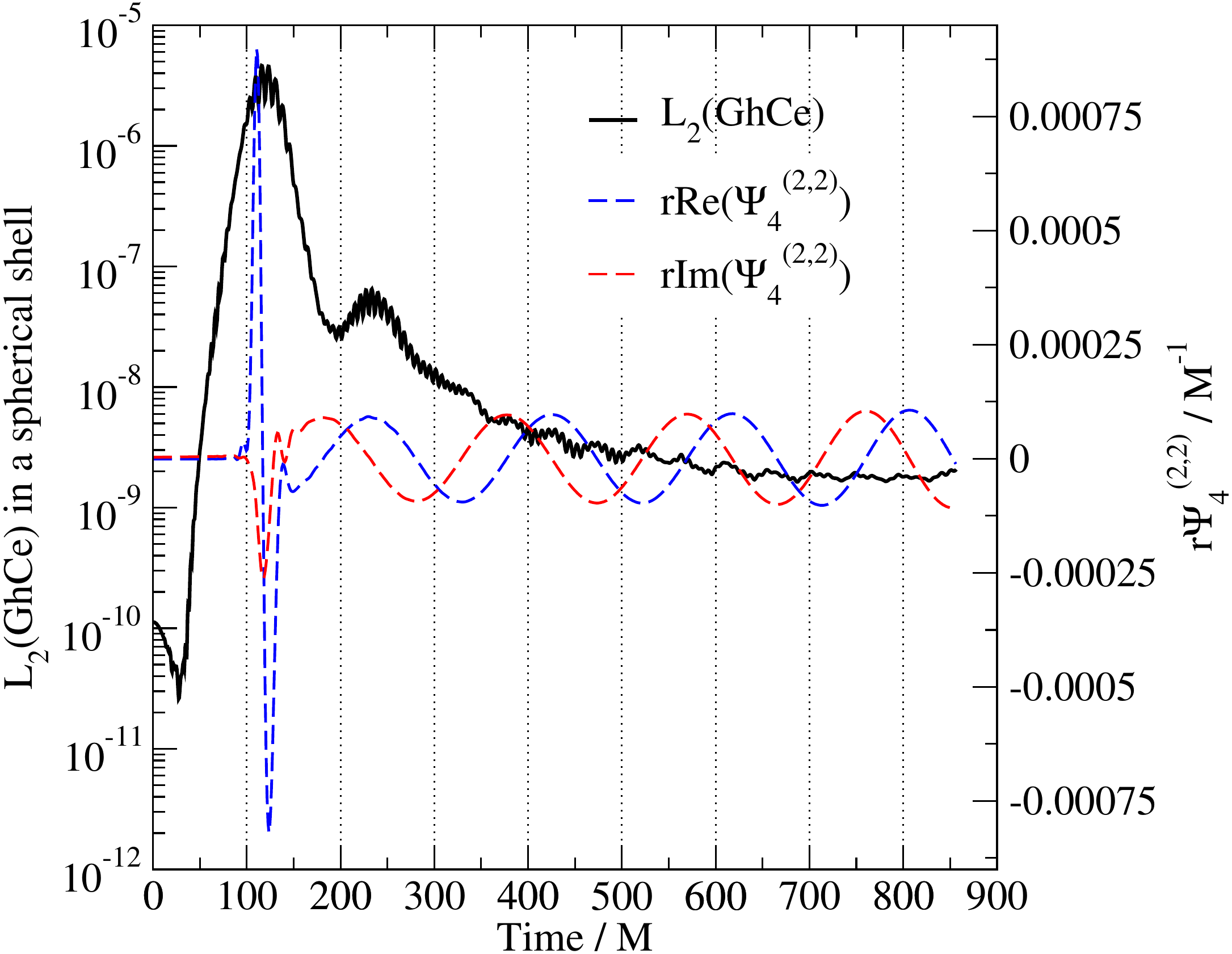}
\caption{The constraint violation and waveform near the start of the standard
  example simulation. 
  The dashed blue and red lines are $r$ times the real and imaginary
  parts of the $(l=2,m=2)$ mode of $\Psi_4$ wave extracted at a coordinate
  radius of $110$M. The Solid black curve is the $L_2$ norm of the constraint
  energy $\text{GhCe}$ within a spherical shell extending radially from $99$M
  to $120$M (i.e. containing the wave extraction sphere). Comparing these
curves, we see that constraint violation remains elevated for a long time after
the junk radiation (the sharp features at the beginning of the waveforms) has
already passed through the spherical shell.}
\label{fig:JunkOverTime}
\end{figure}

To demonstrate this relative longevity of constraint violation, we plot, in
Fig.~\ref{fig:JunkOverTime}, the $(l=2,m=2)$ mode gravitational wave $r\Psi_4$
for the standard equal-mass non-spinning binary simulation extracted at a sphere
$110$M from the coordinate center, as well as the $L_2$ norm of the constraint
violation in a spherical shell subdomain (extending radially from $99$M to
$120$M) surrounding that extraction radius. The constraint violation is measured
using the Generalized Harmonic Constraint Energy ($\text{GhCe}$) defined
in \cite{Lindblom2006}, which includes modes that violate the generalized
harmonic gauge constraints, and those secondary constraints introduced when
reducing the evolution equations to first order. The $L_2$ norm is defined as 
\bea
L_2(\text{GhCe})= \sqrt{\sum_{q=1}^N \frac{\text{GhCe}({\bf x}_q)^2}{N}}
\eea
where ${\bf x}_q$ are the spectral collocation points and $N$ their total population. 

From Fig.~\ref{fig:JunkOverTime}, we see that the junk radiation may excite the
constraint violation by many orders of magnitude, which remains elevated long
after $t\approx 150$M when according to the $r\Psi_4$ curves, one would estimate
the JR to have exited the spherical shell under consideration. Indeed even the
JR itself may have a tail (invisible in $r\Psi_4$ but possibly connected to the
oscillations in $L_2(\text{GhCe})$) resulting from new JR being excited by the
primary one, which can initially travel back towards the origin
\cite{Boyle2007}. A hint of this is shown in Refs.~\cite{Boyle2007,Scheel2009}
where a secondary pulse of JR has been observed to last for two additional light-crossing
times after the primary JR has already exited.

\begin{figure}[!t]
\includegraphics[width=0.85\columnwidth]{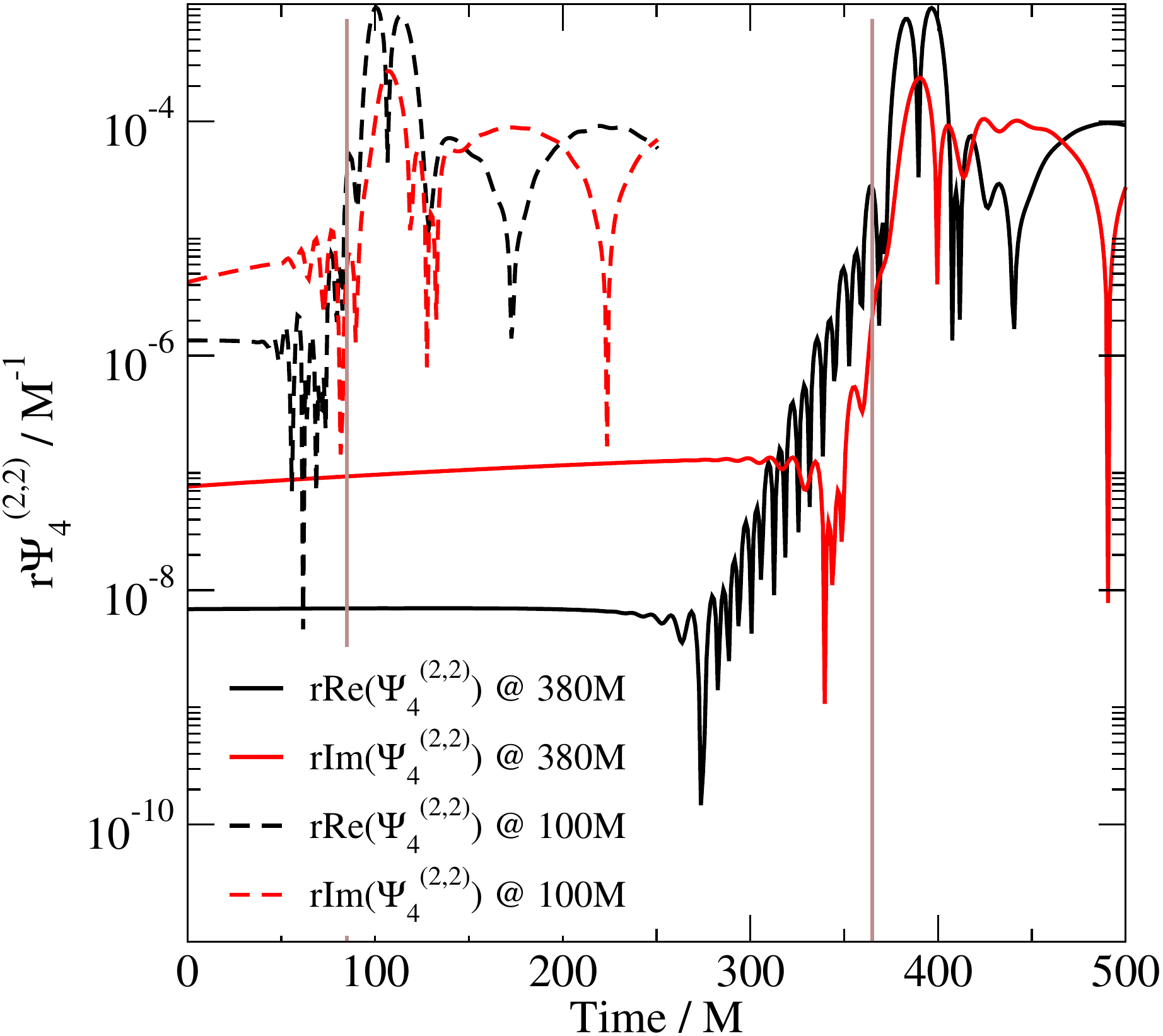}
\caption{The junk radiation in the $r\Psi_4$ waveforms extracted at coordinate radii
$r_{\text{ex}}=100$M and $380$M, denoted by dashed and solid curves
respectively. The expected arrival times, as $r_{\text{ex}}$ minus black hole
initial separation, are plotted as vertical lines. The peak of JR stay
relatively consistent in its time lag from the expected arrival time, but the
superluminally moving leading edge runs further ahead in the case where
$r_{\text{ex}}$ is larger.}
\label{fig:WideningJR}
\end{figure}

Before moving on, we note that the reason constraint violation starts rising
before the expected JR arrival time of $99\text{M}-15\text{M}=84$M is because
the JR tends to widen over time when under-resolved. Namely, upon entering into
a new subdomain, JR corrupts the spectral representation of the entire subdomain
simultaneously, thus appears to teleport instantaneously to the other side of
that subdomain, giving the appearance of a superluminally moving leading edge.
This effect is demonstrated in Fig.~\ref{fig:WideningJR}.

\subsection{Impact on eccentricity estimate \label{sec:JROnOE}}
\begin{figure}[!t]
\includegraphics[width=0.99\columnwidth]{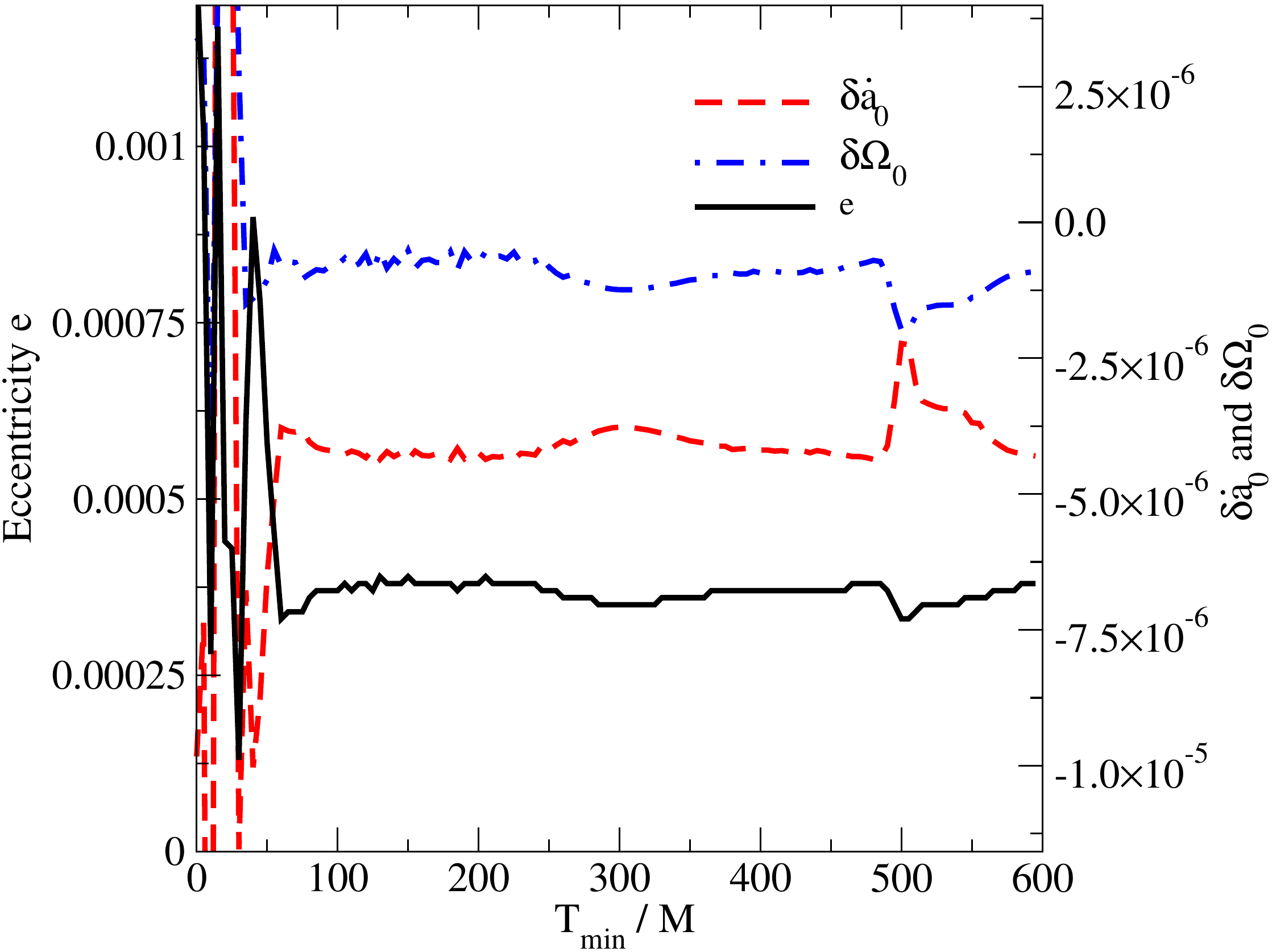}
\caption{The dependence on $T_{\text{min}}$ of eccentricity $e$ (left vertical
  axis) estimated by Eq.~\eqref{eq:Pfeiffer-Brown-etal:2007Estimate} and $(\delta
  \dot{a}_0,\delta \Omega_0)$ (right vertical axis) suggested by
  Eq.~\eqref{eq:Pfeiffer-Brown-etal:2007Deltas}, for the first traditional approach
  eccentricity reduction iteration (see Sec.~\ref{Sec:Result} below) of the standard
  example simulation.   
}
\label{fig:ParaAndEcc}
\end{figure}

Junk radiation complicates the eccentricity removal procedure by introducing
high frequency and large amplitude components into $d\omega/dt$, making fitting
by Eq.~\eqref{eq:Pfeiffer-Brown-etal:2007Fitting} difficult. Therefore one has to wait for
the transient effects created by junk radiation to die down before a fitting can
be made. One does this by specifying a $T_{\text{min}}$, while the maximum time
in the fitting interval $T_{\text{max}}$ is determined accordingly as
$T_{\text{min}} + 5\pi/\Omega_0$, i.e. fitting is done with just over two orbits
after $T_{\text{min}}$. 

The dependence on $T_{\text{min}}$ is shown in Fig.~\ref{fig:ParaAndEcc} for the
standard example simulation. We once again observe a long-lasting impact of JR
that prevents $\delta \Omega_0$ and $\delta \dot{a}_0$ estimates from settling
down quickly. Aside from the ingoing and secondary JR and the lingering constraint violation
(which is less severe for the subdomains close to the black holes that have
higher resolution), we have a new complication in this case. 
 We first recall that although the data being evolved
 corresponds to that measured in an ``inertial frame'' whose coordinates
 $\{x_I\}$ correspond to inertial observers asymptotically, the spectral
 computation is done on a co-moving ``grid'' coordinate system $\{x_G\}$ in
 which black holes remain undeformed and located at fixed positions
 \cite{Scheel2006}. A feedback control system \cite{Hemberger:2012jz} is used to
 connect the two coordinate systems by essentially tracking the motion of the
 black holes in the inertial frame. One parameter in this control system that
 tracks the orbital rotation of the black holes is used as the $\omega$ variable
 appearing in Eq.~\eqref{eq:Pfeiffer-Brown-etal:2007Fitting}. Therefore, the stability of
 $e$, $\delta \Omega_0$ and $\delta \dot{a}_0$ estimates depends on that of the
 feedback control system. However, the response of such a system to the passing
 of a violent disturbance such as JR would not generally die away
 instantaneously, even if the system is ultimately stable. 

\begin{figure}[!t]
\includegraphics[width=0.99\columnwidth]{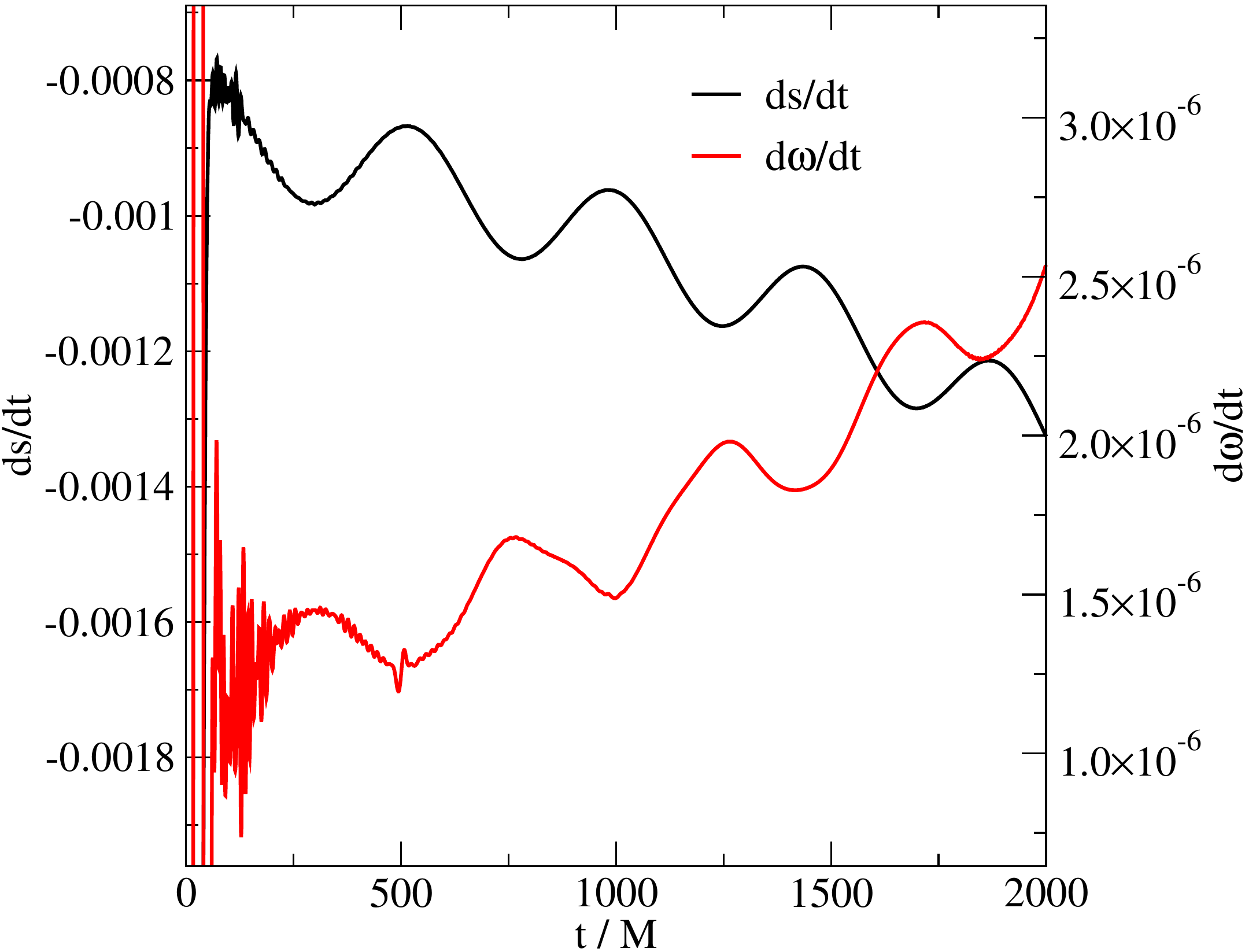}
\caption{Comparison between $ds/dt$ and $d\omega/dt$ for the same simulation as
shown in Fig.~\ref{fig:ParaAndEcc}.}
\label{fig:CompareOmegaWithS}
\end{figure}

To gauge the importance of this effect, we compare the time derivatives of
spatial separation $s$ of the black hole apparent horizons to that of $\omega$.
Because $s$ is calculated as an integral using the spatial metric, it is a
physical quantity that's not directly influenced by the control system. 
We therefore expect $ds/dt$ to settle down much more quickly than $d\omega/dt$.
This is indeed the case as shown by Fig.~\ref{fig:CompareOmegaWithS}. By visual
inspection, it is clear that the deviation from a smooth (drifting) sinusoidal
curve is pronounced in $d\omega/dt$ up to around $t=1500$M. On the other hand,
$ds/dt$ appears to have settled down before $t=500$M. 
In the next section, we introduce a procedure that helps avoid even those high
frequency oscillations at the beginning of $ds/dt$.  

\section{The joint-elimination algorithm \label{sec:Method}} 
We now turn to the joint-elimination method outlined in
Sec.~\ref{sec:Intro}, which as already alluded to in that section, ameliorates
those longer-lasting impacts of JR examined in Sections \ref{Sec:constraints}
and \ref{sec:JROnOE}. 
This section is devoted to a brief description of the technical details of 
how we process the periodically intercepted evolution data and feed it through
the initial data solver. 

For the purpose of this paper, the only adjustments to the intercepted data are
those required by eccentricity reduction, and details of our implementation of
these adjustments in the stop-and-go operation differs significantly from the
traditional approach described in Sec.~\ref{Sec:EcRed}.  
First, the changes to the black holes' radial and tangential velocities
are now applied at the time of stopping $t=T_{\text{stop}}$ instead of initial
time $t=0$, so the phase $\phi$ in Eq.~\eqref{eq:Pfeiffer-Brown-etal:2007Deltas} should be
changed to that at $T_{\text{stop}}$. This can be accomplished by substituting
$t$ in Eq.~\eqref{eq:Pfeiffer-Brown-etal:2007Fitting} with $t-T_{\text{stop}}$. Migrating to
$T_{\text{stop}}$ affords us some flexibility. For example, we can now choose
$T_{\text{stop}}$ such that $\phi=0,\pi$ or $\phi=\pi/2,3\pi/2$ so that only one
of $\delta \Omega_{\text{stop}}$ and $\delta \dot{a}_{\text{stop}}$ is
non-vanishing. For the simple example case we look at in Sec.~\ref{Sec:Result},
there doesn't appear to be any benefit in doing so, yet this option may become
useful in more complicated binary configurations. 

Given $(\delta \Omega_{\text{stop}}, \delta \dot{a}_{\text{stop}})$ from
Eq.~\eqref{eq:Pfeiffer-Brown-etal:2007Deltas} with the appropriate $\phi$, we now turn to
the problem of implementing them. Because the mean curvature for the intercepted
data is no longer vanishing, the dual frame procedure used in
Sec.~\ref{Sec:IDSum} is no longer applicable, and we solve the XCTS equations
directly in the inertial frame instead. In any case we have realistic
$\partial_tK$ and $\tilde{u}_{ij}$, so there is no need to invoke the
co-rotating frame. To implement $(\delta \Omega_{\text{stop}}, \delta
\dot{a}_{\text{stop}})$, we simply add them to the $\xi_{\rm rad}^i$ term that
appears in the inertial frame inner boundary condition
\eqref{eq:BCInertialFrame}, which now becomes 
\bea
\xi^i_{\rm rad} = \left(({\bf \Omega}_{\text{stop}} + \delta {\bf
\Omega}_{\text{stop}} )\times {\bf r}\right)^i + \left(\dot{a}_{\text{stop}} +
\delta \dot{a}_{\text{stop}}\right) r^i.
\eea
Because the elliptic solver is not allowed to change boundary conditions, we do
not need to worry that the solution to the XCTS equations simply revert back to the
original intercepted data.

The simplest way to impose this new shift inner boundary condition is to weigh
$(\delta {\bf \Omega}_{\text{stop}} \times {\bf r})^i + \delta
\dot{a}_{\text{stop}} r^i$ by Gaussian envelops like those in
Eq.~\eqref{eq:SKSFreeData}, add the result onto the intercepted shift, and then
extract the Dirichlet boundary condition from this augmented shift. This is done at
both the inner (excision surfaces) and outer (the outer edge of the
computational domain instead of $r\rightarrow \infty$) boundaries.  
Dirichlet boundary conditions using intercepted lapse is used for $N$, while
uniform constant $1$ is used for $\psi$. 
The augmented shift itself provides a smooth (no discontinuity at the
boundaries) initial guess of $N^i$ for the elliptic solver, while intercepted
$N$ and constant field $\psi=1$ complete the rest of the initial guesses. 

In addition to physical data, many auxiliary quantities are needed during an
evolution using the \verb!SpEC! code, such as the parameters used by the
feedback control system mentioned in Sec.~\ref{sec:JROnOE}. Because these
parameters don't change physics, we simply keep their value unchanged through
the stop-and-go operation, even though the physical data has been altered
slightly. This causes small oscillations in the parameters immediately after
relaunch, which does not signal the existence of a new physical junk radiation
and tends to settle down relatively quickly for small $(\delta
\Omega_{\text{stop}}, \delta \dot{a}_{\text{stop}})$ (See
Fig.~\ref{fig:EccReductionCompare} for an example). 

Another complication is that although the XCTS equations are solved with a
multidomain spectral method \cite{Pfeiffer2003}, its current preferred domain
decomposition 
is different from that of the time evolution \cite{Szilagyi:2009qz}, 
due partially to historical reasons. A consequence of involving all these
different numerical grids is that filtering on the spectral coefficients
\cite{Szilagyi:2009qz} is required to avoid aliasing effects when copying data
between them. 

\section{The results \label{Sec:Result}}
In this section, we demonstrate that the three issues targeted by our
joint-elimination procedure, namely junk radiation, orbital eccentricity and
constraint violation, are resolved as expected. We do so through illustration
using the standard example simulation. For comparison, we also carry out the
traditional eccentricity reduction procedure of repeatedly starting the simulation
from $t=0$ using initial states constructed with analytical free-data. 
We will refer to this alternative as the ``traditional approach''. 

\subsection{Junk radiation reduction \label{Sec:JRReduction}}
\begin{figure}[!t]
\includegraphics[width=0.99\columnwidth]{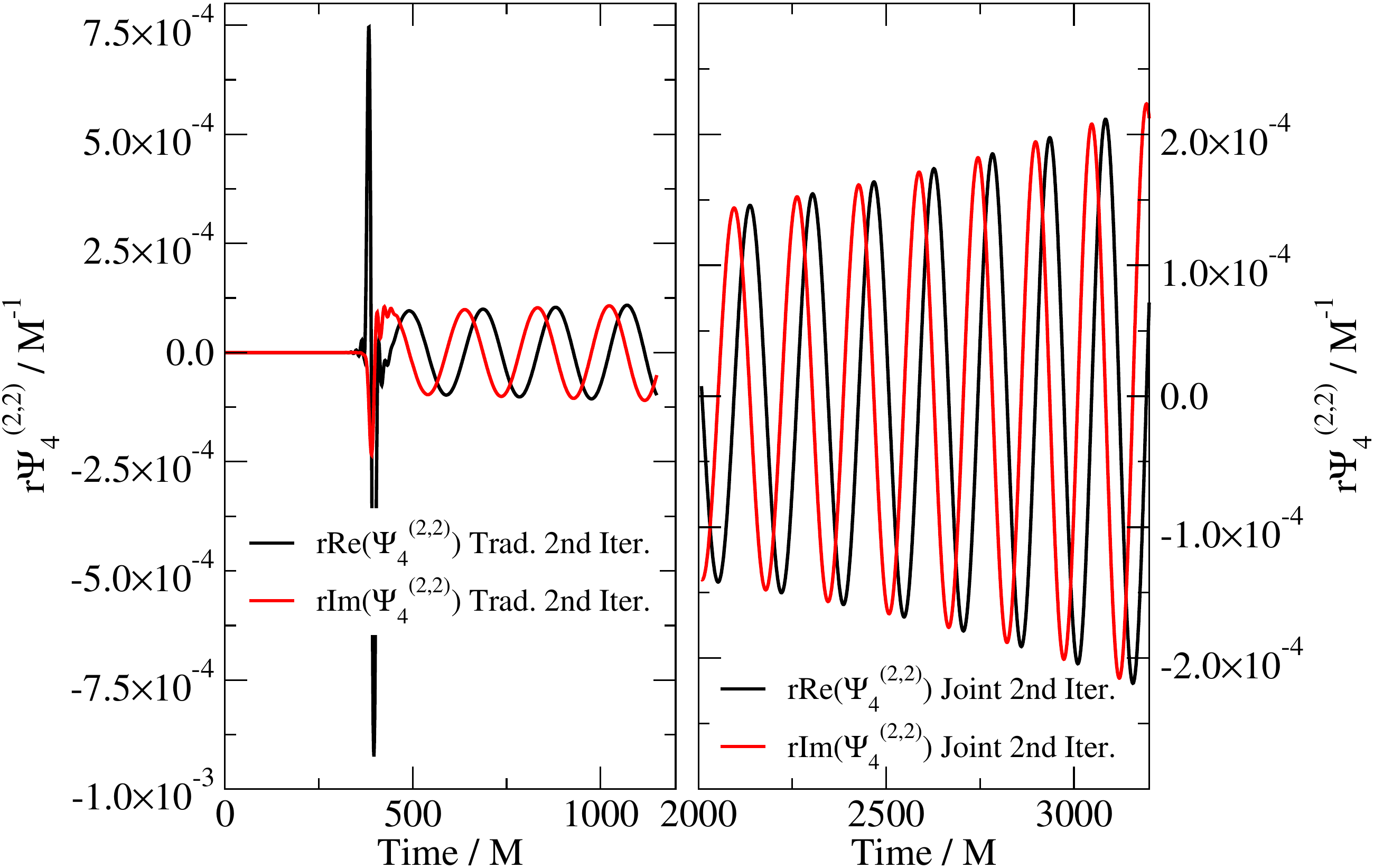}
\caption{The $(l=2,m=2)$ mode $r\Psi_4$ waveforms extracted at coordinate radius
of $380$M, from the evolutions using initial data generated with the
traditional (left panel) and joint-elimination (right panel) approaches. 
}
\label{fig:Waveform}
\end{figure}

\begin{figure}[!b]
\includegraphics[width=0.99\columnwidth]{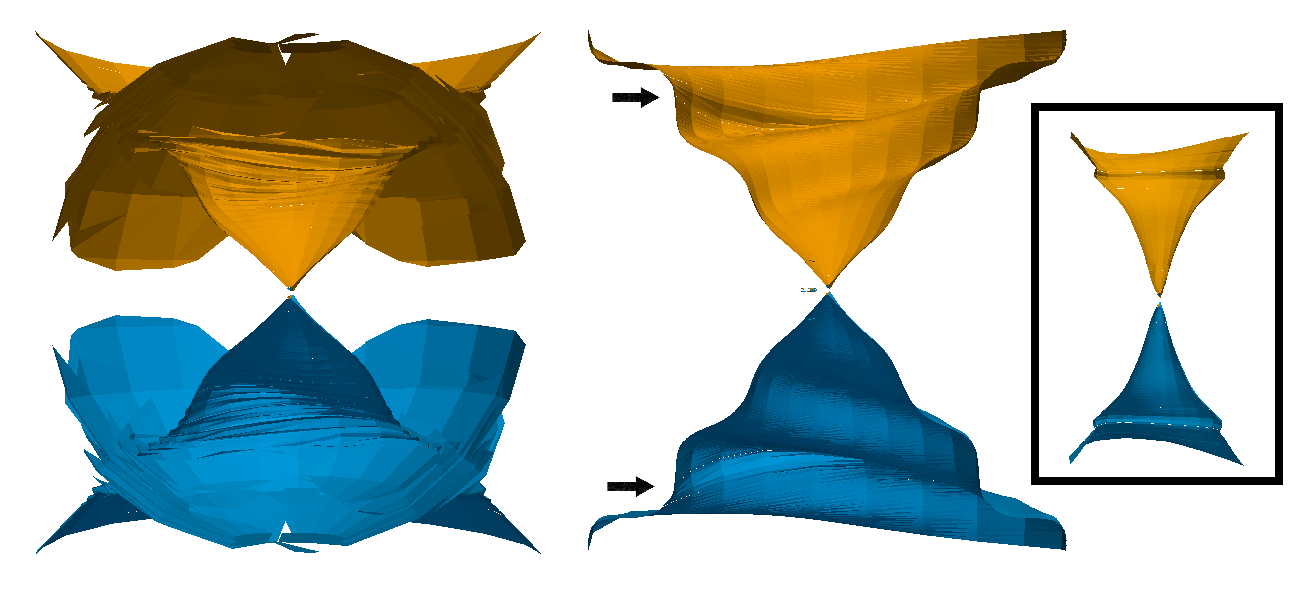}
\caption{Left: Two contours of $\theta^t$ in the simulation starting from the
traditional approach initial data, after eccentricity has been removed. There is
an obvious junk pulse moving outwards. Right: Two contours of $\theta^t$ in the
simulation starting from the joint-elimination approach initial data. The only
visible new junk radiation is a pulse of small disturbance moving from the outer
boundary inwards, as highlighted by the arrows. The inset in the black frame
shows a continuation of a clean simulation (after initial JR has exited), but with
outer boundary shifted inwards. A new pulse of JR resulting from the mismatch
between bulk data and evolution boundary condition can be seen moving inwards. }
\label{fig:InwardBdry}
\end{figure}

Although we are assured, 
by construction, the primary junk
radiation should have passed by the time we are done with the eccentricity
reduction iterations, we nevertheless need to make sure that no significant
amount of new junk radiation is being introduced by the stop-and-go operation
discussed in Sec.~\ref{sec:Method}.  Fig.~\ref{fig:Waveform} compares the
$(l=2,m=2)$ mode waveforms in the final iteration of the traditional and
joint-elimination approaches. It confirms that there is no visible junk
radiation in the joint-elimination approach waveform. 

Next, we move on to the $\theta^t$ contours, which are more sensitive JR
detectors. These contours are shown in Fig.~\ref{fig:InwardBdry}, with the
result for traditional approach on the left and joint-elimination approach on
the right. The only visible new JR is an incoming pulse from the outer boundary
(highlighted by black arrows), which is due to the inconsistency between the
outer boundary condition imposed by the evolution system and the initial data
inside the computational domain. 
Specifically, we use the constraint preserving boundary condition (Eq.~(95) of
\cite{Buchman2006}) that sets the incoming gravitational radiation to zero, while in
general such radiation does exist at finite radius. During a continuous evolution, 
the persistent imposition of the boundary condition
forces the incoming radiation to vanish at the outer boundary (which is part of
the reason why outer boundary has to be placed far away). However, after we
change the black hole velocities and copy data between different grids
(introducing aliasing noise), this is no longer exactly true when we relaunch 
the simulation. The impact of this
mismatch can be seen more cleanly if instead of carrying out our stop-and-go
operation, we simply restart a stopped simulation, but move the outer boundary
inwards to where incoming radiation is still present. The $\theta^t$ contours
in this case are plotted in the inset of Fig.~\ref{fig:InwardBdry}, and we see a
similar incoming pulse. This subtle effect is invisible in the waveform shown in
Fig.~\ref{fig:Waveform}.  

\subsection{Orbital eccentricity reduction \label{Sec:OEReduction}}
\begin{figure}[!t]
\includegraphics[width=0.85\columnwidth]{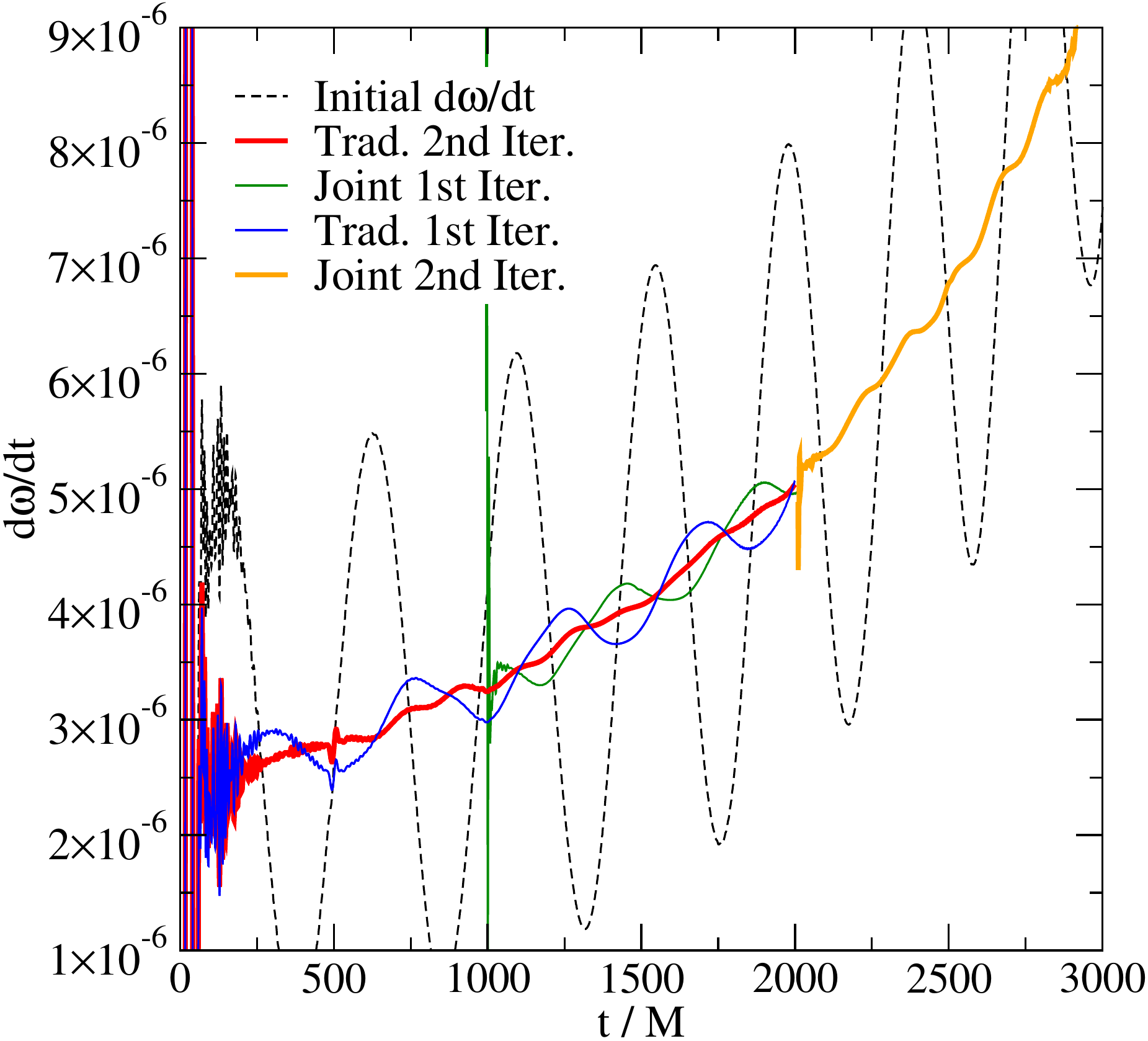}
\caption{The $d\omega/dt$ curves for the initial and subsequent eccentricity
  reduction iterations, with both the traditional approach and the 
  joint-elimination approach shown. With both approaches, the oscillations in
  $d\omega/dt$ become dominated by frequencies higher than the orbital one during 
  the second iteration. 
}
\label{fig:EccReductionCompare}
\end{figure}

We turn next to the issue of eccentricity reduction. With both the traditional
and joint-elimination approaches, we manage to reduce eccentricity from around
$0.0033$ to under $0.0001$ within two iterations. With the traditional approach, 
the sequence of eccentricities
is $0.0033 \rightarrow 0.00037 \rightarrow 0.00009$, 
and in the joint-elimination approach the sequence is
$0.0033 \rightarrow 0.00025 \rightarrow 0.00005$. 
Below an eccentricity of $10^{-4}$, the oscillation in $d\omega/dt$
becomes dominated by frequencies higher than orbital
frequency (see the red and orange lines in Fig.~\ref{fig:EccReductionCompare}),
signaling a possibly non-eccentricity related origin \cite{Buonanno:2010yk}, 
and the fitting formula \eqref{eq:Pfeiffer-Brown-etal:2007Fitting} becomes ill-behaved. 

\begin{figure}[!t]
\includegraphics[width=0.95\columnwidth]{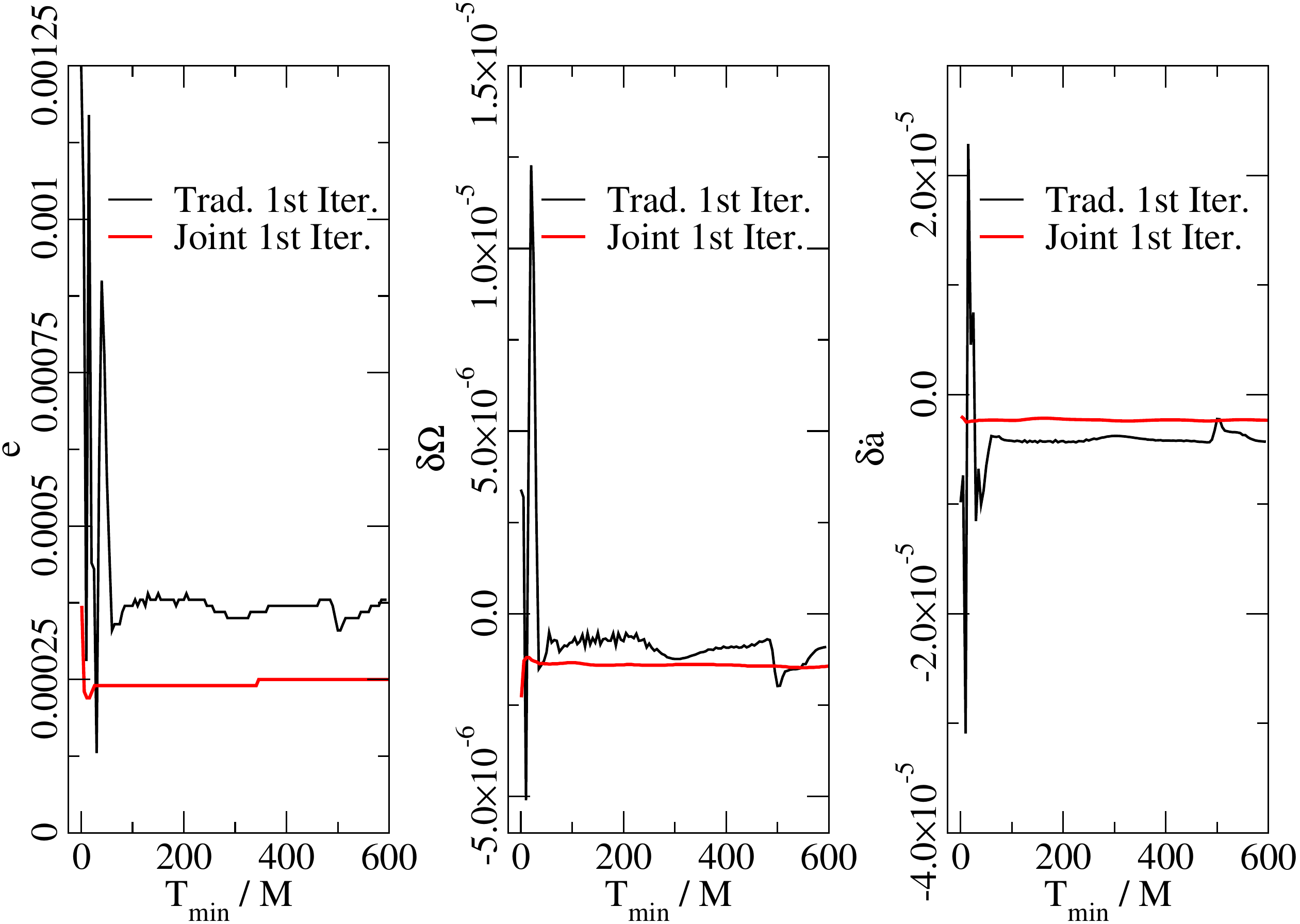}
\caption{
  Comparison between the dependence on $T_{\text{min}}$ of $e$ (left
panel), $\delta \Omega$ (middle panel) and $\delta \dot{a}$ (right panel) for the
traditional and joint-elimination approaches. These parameters are extracted
during the first eccentricity reduction iteration. For better visual comparison
between curves, the time axis for the joint-elimination approach has been
shifted such that it appears to start at $t=0$. The simulations in both
approaches are evolved for a sufficiently long time to accommodate
$T_{\text{max}}$ for all $T_{\text{min}}$ used in this figure. }
\label{fig:ParaAndEccResult}
\end{figure}

Although for this simple example, the final eccentricity value is only marginally smaller 
in the joint-elimination approach as compared to the traditional approach, 
the improvement in the stability (against
$T_{\text{min}}$) of estimation for $e$, and subsequently $\delta \Omega$ and
$\delta \dot{a}$ (we drop the subscripts for brevity), is appreciable 
in the joint-elimination approach, making
the eccentricity reduction process more robust. To demonstrate this stability, we plot $e$,
$\delta \Omega$ and $\delta \dot{a}$ against $T_{\text{min}}$ in
Fig.~\ref{fig:ParaAndEccResult}, obtained during the first iteration in both the
traditional and joint-elimination approaches. Aside from the initial disturbance
caused by JR directly, the quality of parameter estimation in the traditional
approach continues to suffer for a long period afterwards, while the situation
is improved with the joint-elimination approach. 

\subsection{Constraint violation reduction \label{Sec:CVReduction}}
\begin{figure}[!t]
\includegraphics[width=0.85\columnwidth]{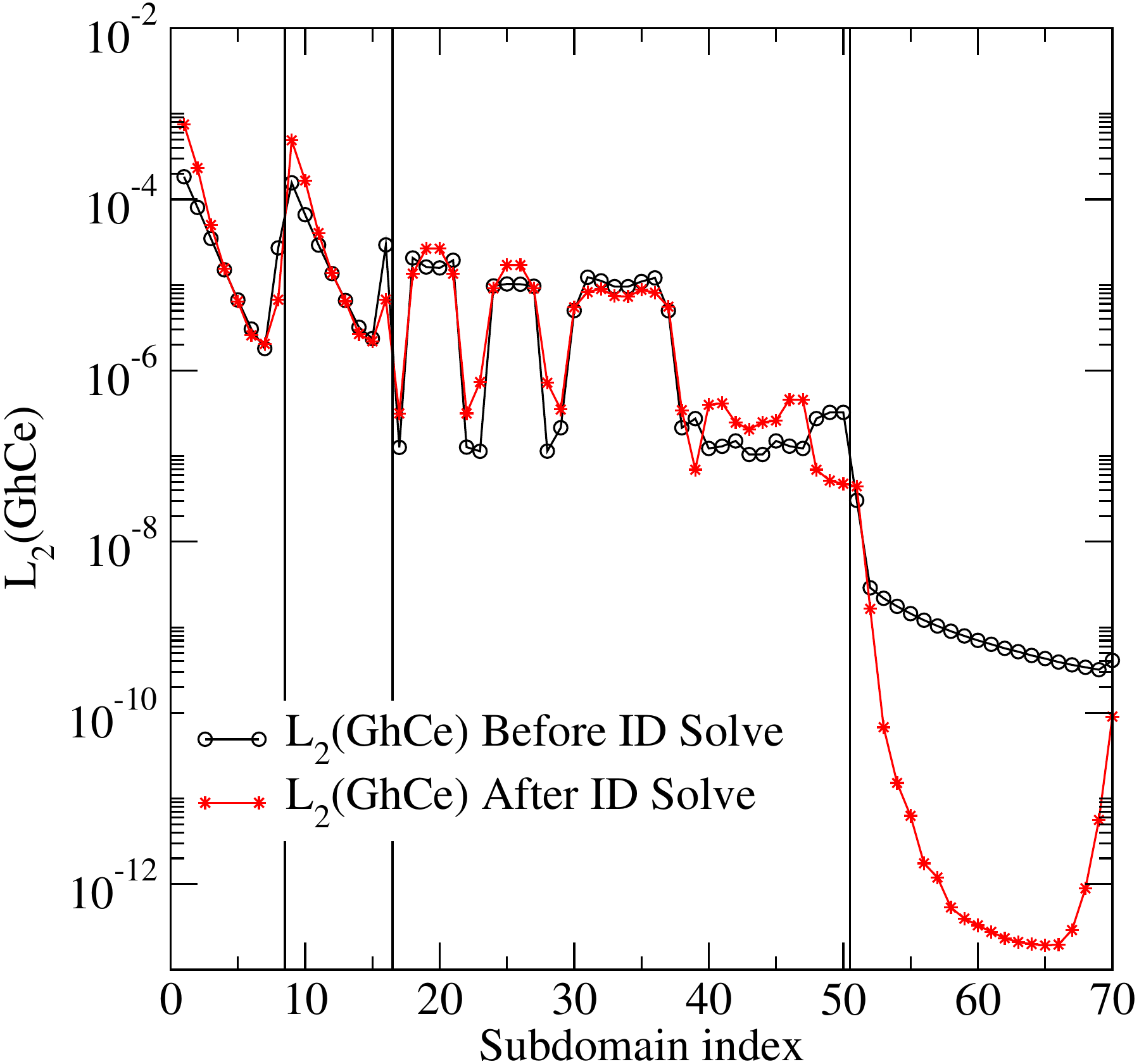}
\caption{
  The $L_2$ norm of the $\text{GhCe}$ constraint measurement within
  individual subdomains, calculated at the beginning of the first iteration 
in the joint-elimination approach. For comparison, we also show the constraint violation 
just before the stop-and-go operation, i.e. before the initial data solver is 
invoked. 
The horizontal axis is the index for the subdomains.
The sectors separated by vertical
lines contain, from left to right, inner spherical shells around one of the
black holes extending from the excision boundary to $4.5$M, similar shells
around the other black hole, some cylindrical shapes filling up the region
in-between and surrounding the inner spherical shells, and finally outer
spherical shells extending from $45$M to $480$M that enclose all of the
aforementioned subdomains. 
}
\label{fig:ConstraintAfterStop}
\end{figure} 

Finally, we examine the consequence of re-imposing constraints when we
periodically feed the data through the initial data solver, which by construction
solves for the constraint parts of the Bianchi identity. Furthermore, when we
assemble the output of the initial data solver into variables used by the
generalized harmonic evolution system \cite{Lindblom:2007}, we also re-impose the
secondary constraints coming from reducing a second order partial differential
equation into a system of first order equations. For example, one of the
evolution variables is $\kappa_{abc}$, which should relate to spacetime metric
${}^{(4)}{\bf g}$ by 
\bea
\kappa_{0ab}&=&-\frac{1}{N}\left( \partial_t {}^{(4)}g_{ab} - N^i \partial_i
{}^{(4)}g_{ab} \right), \notag \\ 
\kappa_{iab} &=& \partial_i {}^{(4)}g_{ab}.
\eea
This relationship is reaffirmed when we reconstruct $\kappa$ from the spatial
and temporal derivatives of the metric. 
These attributes of the stop-and-go operation would help clean up the JR-excited
constraint violation discussed in Sec.~\ref{Sec:constraints}.

In this subsection, we verify that the effectiveness of the initial data solver  
is not compromised by the complexity of numerical 
metric data, 
or our black hole velocity adjustments. In other words, the constraint violation
is reduced as expected after an iteration of the stop-and-go operation. 
To this end, we compare the constraint violation at the beginning ($t=994.5$M) 
of the first iteration in the joint-elimination approach,
to that just before the stop-and-go operation. 
We plot in Fig.~\ref{fig:ConstraintAfterStop} the Generalized 
Harmonic Constraint Energy ($\text{GhCe}$) for different subdomains in these two 
cases.   
The
points in the right-most sector are associated with spherical shell subdomains
situated on the outside of the entire computational domain. Because the grid structure 
in these regions are least appropriate for resolving JR, 
we expect the reduction in constraint violation 
to be most pronounced for them, and this is exactly what we observe.   

\begin{figure}[!t]
\includegraphics[width=0.95\columnwidth]{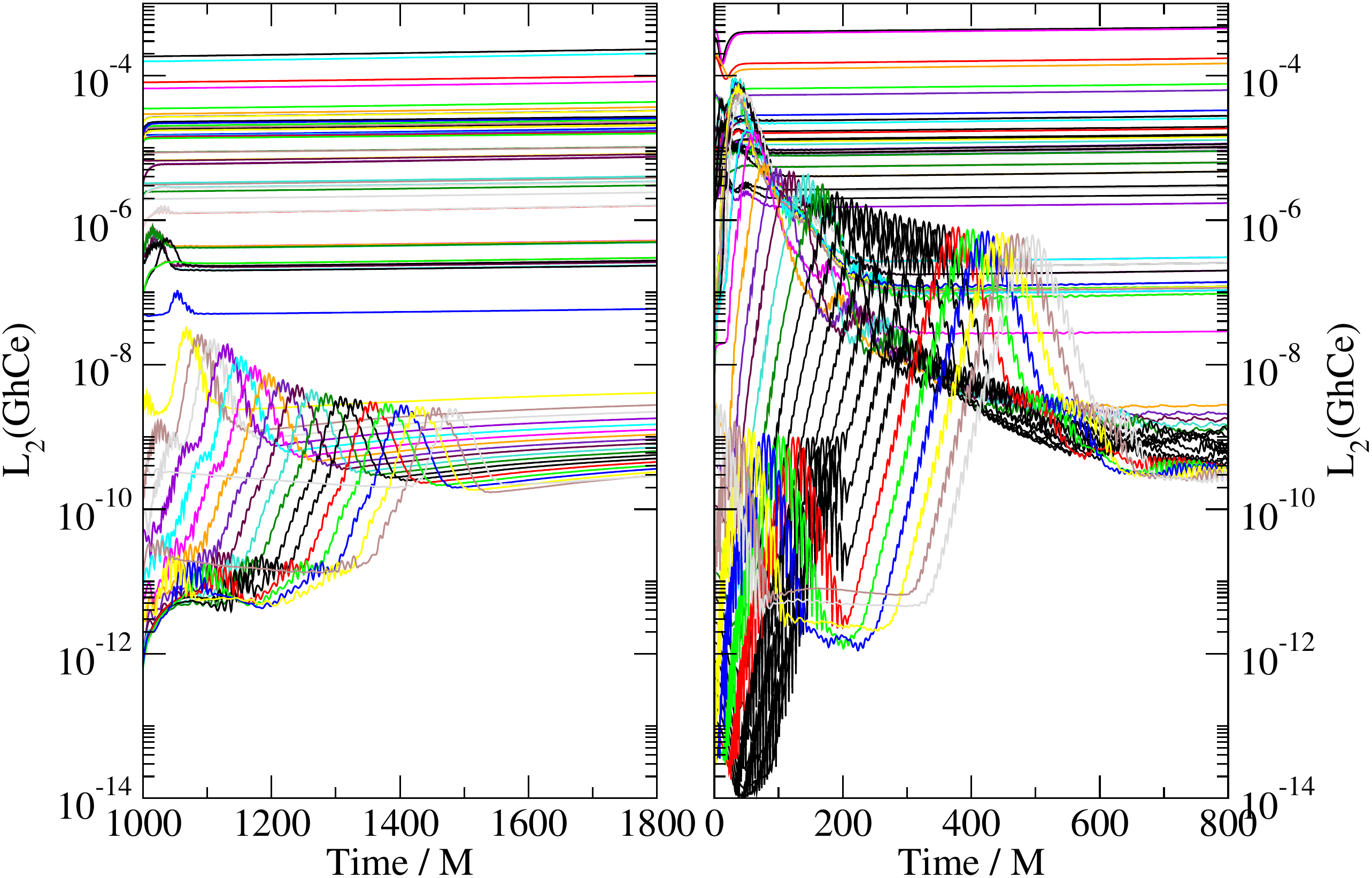}
\caption{ The evolution of $L_2(\text{GhCe})$ broken down to individual subdomains, for the
joint-elimination (left panel) and traditional (right panel) approaches. In both cases, the
constraint violation asymptotes to the levels sustainable by the (shared)
resolution of the simulations. During the intermediate time, constraint violation is smaller
for the joint-elimination case due to the reduction in JR. 
The curves near the bottom of the plots correspond to outer
spherical shell subdomains.
}
\label{fig:GrowthOfConstraint}
\end{figure}

When we launch the numerical evolution from these initial data, the constraint violation
will rise towards the levels sustainable by our low simulation resolution. One may
of course increase the resolution and/or adopt a more sophisticated domain decomposition to
take better advantage of the clean initial data. However, even for low resolution, 
the constraint violation behaves much less aggressively during the intermediate time,
due to the reduction in JR.
This is graphically depicted in Fig.~\ref{fig:GrowthOfConstraint}, in which we plot the
$L_2(\text{GhCe})$ growth per subdomain
for the joint-elimination and traditional approaches, during the first eccentricity reduction
iteration.

\section{Conclusion}
In this paper we have argued that the constraint violation excited by 
under-resolved junk radiation would be long-lived, motivating the need to remove
JR's impacts with a method better than simply waiting for it to exit. 
We also propose that the robustness of the eccentricity reduction procedure
could be improved if we do not have to contend with JR. 
We then introduced a practical method for achieving these goals.
In short, we link up the iterations in an eccentricity reduction procedure by feeding the 
final numerical metric data in the previous iteration into the initial data solver for the 
next. The effectiveness 
and advantages of our approach is then demonstrated with a particular binary example. 

Aside from generating constraint violation and disturbing control systems,
another consequence of JR is that it alters the physical parameters of the
black holes. For example the spins of the black holes may drop under the
influence of JR, which hampers efforts at creating initial data for high-spin
black hole binaries. The joint-elimination approach introduced here can
be adapted to address this problem. Namely we could change the $N^i_{\parallel}$
term in the inertial frame boundary condition Eq.~\eqref{eq:BCInertialFrame}
during each iteration, in order to dial up the black hole spins in stages.
Similar arrangements can be made for an adjustment of the black holes' masses.  
Applications such as this will be the subject of
further studies. 

\acknowledgments
We would like to thank Nicholas Taylor and Mark Scheel for helpful discussions,
Harald Pfeiffer, William Throwe, Saul Teukolsky and Daniel Hemberger
for reading a draft of the
paper and providing valuable comments.  
This research is supported by NSF grants 
PHY-1068881 and PHY-1005655, NASA grants NNX09AF97G and NNX09AF96G,
and by the Sherman Fairchild Foundation and the Brinson Foundation.
The numerical work was carried out on the Caltech computer cluster \textsc{zwicky}, 
funded by the Sherman Fairchild Foundation and the NSF MRI-R2 grant PHY-0960291.

\bibliography{JunkRadiationPaper.bbl}

\begin{thebibliography}{48}
\expandafter\ifx\csname natexlab\endcsname\relax\def\natexlab#1{#1}\fi
\expandafter\ifx\csname bibnamefont\endcsname\relax
  \def\bibnamefont#1{#1}\fi
\expandafter\ifx\csname bibfnamefont\endcsname\relax
  \def\bibfnamefont#1{#1}\fi
\expandafter\ifx\csname citenamefont\endcsname\relax
  \def\citenamefont#1{#1}\fi
\expandafter\ifx\csname url\endcsname\relax
  \def\url#1{\texttt{#1}}\fi
\expandafter\ifx\csname urlprefix\endcsname\relax\def\urlprefix{URL }\fi
\providecommand{\bibinfo}[2]{#2}
\providecommand{\eprint}[2][]{\url{#2}}

\bibitem[{\citenamefont{{Harry (for the LIGO Scientific
  Collaboration)}}(2010)}]{Harry2010}
\bibinfo{author}{\bibfnamefont{G.~M.} \bibnamefont{{Harry (for the LIGO
  Scientific Collaboration)}}}, \bibinfo{journal}{Class.\ Quantum Grav.}
  \textbf{\bibinfo{volume}{27}}, \bibinfo{pages}{084006}
  (\bibinfo{year}{2010}).

\bibitem[{\citenamefont{{The Virgo Collaboration}}(2009)}]{aVIRGO}
\bibinfo{author}{\bibnamefont{{The Virgo Collaboration}}},
  \emph{\bibinfo{title}{{Advanced Virgo Baseline Design}}}
  (\bibinfo{year}{2009}), \bibinfo{note}{{[VIR-0027A-09]}}.

\bibitem[{\citenamefont{{The Virgo Collaboration}}(2012)}]{aVIRGO:2012}
\bibinfo{author}{\bibnamefont{{The Virgo Collaboration}}},
  \emph{\bibinfo{title}{{Advanced Virgo Technical Design Report}}}
  (\bibinfo{year}{2012}), \bibinfo{note}{{[VIR-0128A-12]}},
  \urlprefix\url{https://tds.ego-gw.it/ql/?c=6940}.

\bibitem[{\citenamefont{Somiya}(2012)}]{Somiya:2011np}
\bibinfo{author}{\bibfnamefont{K.}~\bibnamefont{Somiya}}
  (\bibinfo{collaboration}{KAGRA Collaboration}), \bibinfo{journal}{Class.\
  Quantum Grav.} \textbf{\bibinfo{volume}{29}}, \bibinfo{pages}{124007}
  (\bibinfo{year}{2012}), \eprint{1111.7185}.

\bibitem[{\citenamefont{Bishop et~al.}(2011)\citenamefont{Bishop, Pollney, and
  Reisswig}}]{Bishop:2011iu}
\bibinfo{author}{\bibfnamefont{N.}~\bibnamefont{Bishop}},
  \bibinfo{author}{\bibfnamefont{D.}~\bibnamefont{Pollney}}, \bibnamefont{and}
  \bibinfo{author}{\bibfnamefont{C.}~\bibnamefont{Reisswig}},
  \bibinfo{journal}{Class.Quant.Grav.} \textbf{\bibinfo{volume}{28}},
  \bibinfo{pages}{155019} (\bibinfo{year}{2011}), \eprint{1101.5492}.

\bibitem[{\citenamefont{Lovelace}(2009)}]{Lovelace2009}
\bibinfo{author}{\bibfnamefont{G.}~\bibnamefont{Lovelace}},
  \bibinfo{journal}{Class.\ Quantum Grav.} \textbf{\bibinfo{volume}{26}},
  \bibinfo{pages}{114002} (\bibinfo{year}{2009}).

\bibitem[{\citenamefont{Boyle et~al.}(2007)\citenamefont{Boyle, Brown, Kidder,
  Mrou{\'e}, Pfeiffer, Scheel, Cook, and Teukolsky}}]{Boyle2007}
\bibinfo{author}{\bibfnamefont{M.}~\bibnamefont{Boyle}},
  \bibinfo{author}{\bibfnamefont{D.~A.} \bibnamefont{Brown}},
  \bibinfo{author}{\bibfnamefont{L.~E.} \bibnamefont{Kidder}},
  \bibinfo{author}{\bibfnamefont{A.~H.} \bibnamefont{Mrou{\'e}}},
  \bibinfo{author}{\bibfnamefont{H.~P.} \bibnamefont{Pfeiffer}},
  \bibinfo{author}{\bibfnamefont{M.~A.} \bibnamefont{Scheel}},
  \bibinfo{author}{\bibfnamefont{G.~B.} \bibnamefont{Cook}}, \bibnamefont{and}
  \bibinfo{author}{\bibfnamefont{S.~A.} \bibnamefont{Teukolsky}},
  \bibinfo{journal}{Phys.\ Rev.\ D} \textbf{\bibinfo{volume}{76}},
  \bibinfo{eid}{124038} (pages~\bibinfo{numpages}{31}) (\bibinfo{year}{2007}).

\bibitem[{\citenamefont{Peters and Mathews}(1963)}]{PetersMathews1963}
\bibinfo{author}{\bibfnamefont{P.~C.} \bibnamefont{Peters}} \bibnamefont{and}
  \bibinfo{author}{\bibfnamefont{J.}~\bibnamefont{Mathews}},
  \bibinfo{journal}{Phys. Rev.} \textbf{\bibinfo{volume}{131}},
  \bibinfo{pages}{435} (\bibinfo{year}{1963}),
  \urlprefix\url{http://link.aps.org/abstract/PR/v131/p435}.

\bibitem[{\citenamefont{Peters}(1964)}]{Peters1964}
\bibinfo{author}{\bibfnamefont{P.~C.} \bibnamefont{Peters}},
  \bibinfo{journal}{Phys. Rev.} \textbf{\bibinfo{volume}{136}},
  \bibinfo{pages}{B1224} (\bibinfo{year}{1964}),
  \urlprefix\url{http://link.aps.org/abstract/PR/v136/pB1224}.

\bibitem[{\citenamefont{Postnov and Yungelson}(2005)}]{Postnov:2007jv}
\bibinfo{author}{\bibfnamefont{K.}~\bibnamefont{Postnov}} \bibnamefont{and}
  \bibinfo{author}{\bibfnamefont{L.}~\bibnamefont{Yungelson}},
  \bibinfo{journal}{Living Rev. Rel.} \textbf{\bibinfo{volume}{9}},
  \bibinfo{pages}{6} (\bibinfo{year}{2005}), \eprint{astro-ph/0701059}.

\bibitem[{\citenamefont{East et~al.}(2013)\citenamefont{East, McWilliams,
  Levin, and Pretorius}}]{East:2012xq}
\bibinfo{author}{\bibfnamefont{W.~E.} \bibnamefont{East}},
  \bibinfo{author}{\bibfnamefont{S.~T.} \bibnamefont{McWilliams}},
  \bibinfo{author}{\bibfnamefont{J.}~\bibnamefont{Levin}}, \bibnamefont{and}
  \bibinfo{author}{\bibfnamefont{F.}~\bibnamefont{Pretorius}},
  \bibinfo{journal}{Phys.Rev.} \textbf{\bibinfo{volume}{D87}},
  \bibinfo{pages}{043004} (\bibinfo{year}{2013}), \eprint{1212.0837}.

\bibitem[{\citenamefont{Hannam et~al.}(2008)\citenamefont{Hannam, Husa,
  Gonz{\'a}lez, Sperhake, and Br{\"u}gmann}}]{Hannam2007}
\bibinfo{author}{\bibfnamefont{M.}~\bibnamefont{Hannam}},
  \bibinfo{author}{\bibfnamefont{S.}~\bibnamefont{Husa}},
  \bibinfo{author}{\bibfnamefont{J.~A.} \bibnamefont{Gonz{\'a}lez}},
  \bibinfo{author}{\bibfnamefont{U.}~\bibnamefont{Sperhake}}, \bibnamefont{and}
  \bibinfo{author}{\bibfnamefont{B.}~\bibnamefont{Br{\"u}gmann}},
  \bibinfo{journal}{Phys.\ Rev.\ D} \textbf{\bibinfo{volume}{77}},
  \bibinfo{pages}{044020} (\bibinfo{year}{2008}), \eprint{arXiv:0706.1305v2}.

\bibitem[{\citenamefont{Choi et~al.}(2007)\citenamefont{Choi, Kelly, Boggs,
  Baker, Centrella, and van Meter}}]{Choi-Kelly-Boggs-etal:2007}
\bibinfo{author}{\bibfnamefont{D.-I.} \bibnamefont{Choi}},
  \bibinfo{author}{\bibfnamefont{B.~J.} \bibnamefont{Kelly}},
  \bibinfo{author}{\bibfnamefont{W.~D.} \bibnamefont{Boggs}},
  \bibinfo{author}{\bibfnamefont{J.~G.} \bibnamefont{Baker}},
  \bibinfo{author}{\bibfnamefont{J.}~\bibnamefont{Centrella}},
  \bibnamefont{and} \bibinfo{author}{\bibfnamefont{J.}~\bibnamefont{van
  Meter}}, \bibinfo{journal}{Phys.\ Rev.\ D} \textbf{\bibinfo{volume}{76}},
  \bibinfo{pages}{104026} (\bibinfo{year}{2007}).

\bibitem[{\citenamefont{Gonz\'{a}lez et~al.}(2007)\citenamefont{Gonz\'{a}lez,
  Sperhake, Br{\"u}gmann, Hannam, and Husa}}]{Gonzalez2007}
\bibinfo{author}{\bibfnamefont{J.~A.} \bibnamefont{Gonz\'{a}lez}},
  \bibinfo{author}{\bibfnamefont{U.}~\bibnamefont{Sperhake}},
  \bibinfo{author}{\bibfnamefont{B.}~\bibnamefont{Br{\"u}gmann}},
  \bibinfo{author}{\bibfnamefont{M.}~\bibnamefont{Hannam}}, \bibnamefont{and}
  \bibinfo{author}{\bibfnamefont{S.}~\bibnamefont{Husa}},
  \bibinfo{journal}{Phys.\ Rev.\ Lett.} \textbf{\bibinfo{volume}{98}},
  \bibinfo{pages}{091101} (\bibinfo{year}{2007}), \eprint{gr-qc/0610154}.

\bibitem[{\citenamefont{Campanelli et~al.}(2007)\citenamefont{Campanelli,
  Lousto, Zlochower, Krishnan, and Merritt}}]{Campanelli2007b}
\bibinfo{author}{\bibfnamefont{M.}~\bibnamefont{Campanelli}},
  \bibinfo{author}{\bibfnamefont{C.~O.} \bibnamefont{Lousto}},
  \bibinfo{author}{\bibfnamefont{Y.}~\bibnamefont{Zlochower}},
  \bibinfo{author}{\bibfnamefont{B.}~\bibnamefont{Krishnan}}, \bibnamefont{and}
  \bibinfo{author}{\bibfnamefont{D.}~\bibnamefont{Merritt}},
  \bibinfo{journal}{Phys.\ Rev.\ D} \textbf{\bibinfo{volume}{75}},
  \bibinfo{pages}{064030} (\bibinfo{year}{2007}), \eprint{gr-qc/0612076}.

\bibitem[{\citenamefont{Johnson-McDaniel
  et~al.}(2009)\citenamefont{Johnson-McDaniel, Yunes, Tichy, and
  Owen}}]{JohnsonMcDaniel:2009dq}
\bibinfo{author}{\bibfnamefont{N.~K.} \bibnamefont{Johnson-McDaniel}},
  \bibinfo{author}{\bibfnamefont{N.}~\bibnamefont{Yunes}},
  \bibinfo{author}{\bibfnamefont{W.}~\bibnamefont{Tichy}}, \bibnamefont{and}
  \bibinfo{author}{\bibfnamefont{B.~J.} \bibnamefont{Owen}},
  \bibinfo{journal}{Phys.Rev.} \textbf{\bibinfo{volume}{D80}},
  \bibinfo{pages}{124039} (\bibinfo{year}{2009}), \eprint{0907.0891}.

\bibitem[{\citenamefont{Kelly et~al.}(2007)\citenamefont{Kelly, Tichy,
  Campanelli, and Whiting}}]{kellyEtAl:2007}
\bibinfo{author}{\bibfnamefont{B.~J.} \bibnamefont{Kelly}},
  \bibinfo{author}{\bibfnamefont{W.}~\bibnamefont{Tichy}},
  \bibinfo{author}{\bibfnamefont{M.}~\bibnamefont{Campanelli}},
  \bibnamefont{and} \bibinfo{author}{\bibfnamefont{B.~F.}
  \bibnamefont{Whiting}}, \bibinfo{journal}{Phys.\ Rev.\ D}
  \textbf{\bibinfo{volume}{76}}, \bibinfo{pages}{024008}
  (\bibinfo{year}{2007}).

\bibitem[{SXS()}]{SXSWebsite}
  \bibinfo{note}{\url{http://www.black-holes.org/}}.

\bibitem[{\citenamefont{Lindblom
  et~al.}(2006{\natexlab{a}})\citenamefont{Lindblom, Scheel, Kidder, Owen, and
  Rinne}}]{Lindblom:2007}
\bibinfo{author}{\bibfnamefont{L.}~\bibnamefont{Lindblom}},
  \bibinfo{author}{\bibfnamefont{M.~A.} \bibnamefont{Scheel}},
  \bibinfo{author}{\bibfnamefont{L.~E.} \bibnamefont{Kidder}},
  \bibinfo{author}{\bibfnamefont{R.}~\bibnamefont{Owen}}, \bibnamefont{and}
  \bibinfo{author}{\bibfnamefont{O.}~\bibnamefont{Rinne}},
  \bibinfo{journal}{Class.Quant.Grav.} \textbf{\bibinfo{volume}{23}},
  \bibinfo{pages}{S447} (\bibinfo{year}{2006}{\natexlab{a}}),
  \eprint{gr-qc/0512093v3}.

\bibitem[{\citenamefont{Friedrich}(1985)}]{Friedrich1985}
\bibinfo{author}{\bibfnamefont{H.}~\bibnamefont{Friedrich}},
  \bibinfo{journal}{Commun.\ Math.\ Phys.} \textbf{\bibinfo{volume}{100}},
  \bibinfo{pages}{525} (\bibinfo{year}{1985}),
  \urlprefix\url{http://www.springerlink.com/content/w602g633428x8365}.

\bibitem[{\citenamefont{Garfinkle}(2002)}]{Garfinkle2002}
\bibinfo{author}{\bibfnamefont{D.}~\bibnamefont{Garfinkle}},
  \bibinfo{journal}{Phys.\ Rev.\ D} \textbf{\bibinfo{volume}{65}},
  \bibinfo{pages}{044029} (\bibinfo{year}{2002}).

\bibitem[{\citenamefont{Pretorius}(2005)}]{Pretorius2005c}
\bibinfo{author}{\bibfnamefont{F.}~\bibnamefont{Pretorius}},
  \bibinfo{journal}{Class.\ Quantum Grav.} \textbf{\bibinfo{volume}{22}},
  \bibinfo{pages}{425} (\bibinfo{year}{2005}),
  \urlprefix\url{http://stacks.iop.org/0264-9381/22/425}.

\bibitem[{\citenamefont{{M. A. Scheel, M. Boyle, T. Chu, L. E. Kidder, K. D.
  Matthews and H. P. Pfeiffer}}(2009)}]{Scheel2009}
\bibinfo{author}{\bibnamefont{{M. A. Scheel, M. Boyle, T. Chu, L. E. Kidder, K.
  D. Matthews and H. P. Pfeiffer}}}, \bibinfo{journal}{Phys.\ Rev.\ D}
  \textbf{\bibinfo{volume}{79}}, \bibinfo{pages}{024003}
  (\bibinfo{year}{2009}), \eprint{arXiv:gr-qc/0810.1767}.

\bibitem[{\citenamefont{Szilagyi et~al.}(2009)\citenamefont{Szilagyi, Lindblom,
  and Scheel}}]{Szilagyi:2009qz}
\bibinfo{author}{\bibfnamefont{B.}~\bibnamefont{Szilagyi}},
  \bibinfo{author}{\bibfnamefont{L.}~\bibnamefont{Lindblom}}, \bibnamefont{and}
  \bibinfo{author}{\bibfnamefont{M.~A.} \bibnamefont{Scheel}},
  \bibinfo{journal}{Phys.\ Rev.\ D} \textbf{\bibinfo{volume}{80}},
  \bibinfo{pages}{124010} (\bibinfo{year}{2009}), \eprint{0909.3557}.

\bibitem[{\citenamefont{Arnowitt et~al.}(1962)\citenamefont{Arnowitt, Deser,
  and Misner}}]{ADM}
\bibinfo{author}{\bibfnamefont{R.}~\bibnamefont{Arnowitt}},
  \bibinfo{author}{\bibfnamefont{S.}~\bibnamefont{Deser}}, \bibnamefont{and}
  \bibinfo{author}{\bibfnamefont{C.~W.} \bibnamefont{Misner}}, in
  \emph{\bibinfo{booktitle}{Gravitation: An Introduction to Current Research}},
  edited by \bibinfo{editor}{\bibfnamefont{L.}~\bibnamefont{Witten}}
  (\bibinfo{publisher}{Wiley}, \bibinfo{address}{New York},
  \bibinfo{year}{1962}), pp. \bibinfo{pages}{227--265}, \eprint{gr-qc/0405109}.

\bibitem[{\citenamefont{{York, Jr.}}(1979)}]{york79}
\bibinfo{author}{\bibfnamefont{J.~W.} \bibnamefont{{York, Jr.}}}, in
  \emph{\bibinfo{booktitle}{Sources of Gravitational Radiation}}, edited by
  \bibinfo{editor}{\bibfnamefont{L.~L.} \bibnamefont{Smarr}}
  (\bibinfo{publisher}{Cambridge University Press},
  \bibinfo{address}{Cambridge, England}, \bibinfo{year}{1979}), pp.
  \bibinfo{pages}{83--126}.

\bibitem[{\citenamefont{York}(1999)}]{York1999}
\bibinfo{author}{\bibfnamefont{J.~W.} \bibnamefont{York}},
  \bibinfo{journal}{Phys.\ Rev.\ Lett.} \textbf{\bibinfo{volume}{82}},
  \bibinfo{pages}{1350} (\bibinfo{year}{1999}).

\bibitem[{\citenamefont{Pfeiffer and York}(2003)}]{Pfeiffer2003b}
\bibinfo{author}{\bibfnamefont{H.~P.} \bibnamefont{Pfeiffer}} \bibnamefont{and}
  \bibinfo{author}{\bibfnamefont{J.~W.} \bibnamefont{York}},
  \bibinfo{journal}{Phys.\ Rev.\ D} \textbf{\bibinfo{volume}{67}},
  \bibinfo{pages}{044022} (\bibinfo{year}{2003}).

\bibitem[{\citenamefont{Smarr and York}(1978)}]{Smarr78b}
\bibinfo{author}{\bibfnamefont{L.}~\bibnamefont{Smarr}} \bibnamefont{and}
  \bibinfo{author}{\bibfnamefont{J.~W.} \bibnamefont{York}},
  \bibinfo{journal}{Phys.\ Rev.\ D} \textbf{\bibinfo{volume}{17}},
  \bibinfo{pages}{2529} (\bibinfo{year}{1978}).

\bibitem[{\citenamefont{Wilson and Mathews}(1989)}]{Frontiers:Wilson}
\bibinfo{author}{\bibfnamefont{J.~R.} \bibnamefont{Wilson}} \bibnamefont{and}
  \bibinfo{author}{\bibfnamefont{G.~J.} \bibnamefont{Mathews}}, in
  \emph{\bibinfo{booktitle}{Frontiers in Numerical Relativity}}, edited by
  \bibinfo{editor}{\bibfnamefont{C.~R.} \bibnamefont{Evans}},
  \bibinfo{editor}{\bibfnamefont{L.~S.} \bibnamefont{Finn}}, \bibnamefont{and}
  \bibinfo{editor}{\bibfnamefont{D.~W.} \bibnamefont{Hobill}}
  (\bibinfo{publisher}{Cambridge University Press},
  \bibinfo{address}{Cambridge, England}, \bibinfo{year}{1989}), pp.
  \bibinfo{pages}{306--314}.

\bibitem[{\citenamefont{Pfeiffer et~al.}(2003)\citenamefont{Pfeiffer, Kidder,
  Scheel, and Teukolsky}}]{Pfeiffer2003}
\bibinfo{author}{\bibfnamefont{H.~P.} \bibnamefont{Pfeiffer}},
  \bibinfo{author}{\bibfnamefont{L.~E.} \bibnamefont{Kidder}},
  \bibinfo{author}{\bibfnamefont{M.~A.} \bibnamefont{Scheel}},
  \bibnamefont{and} \bibinfo{author}{\bibfnamefont{S.~A.}
  \bibnamefont{Teukolsky}}, \bibinfo{journal}{Comput.\ Phys.\ Commun.}
  \textbf{\bibinfo{volume}{152}}, \bibinfo{pages}{253} (\bibinfo{year}{2003}).

\bibitem[{\citenamefont{Pfeiffer et~al.}(2007)\citenamefont{Pfeiffer, Brown,
  Kidder, Lindblom, Lovelace, and Scheel}}]{Pfeiffer-Brown-etal:2007}
\bibinfo{author}{\bibfnamefont{H.~P.} \bibnamefont{Pfeiffer}},
  \bibinfo{author}{\bibfnamefont{D.~A.} \bibnamefont{Brown}},
  \bibinfo{author}{\bibfnamefont{L.~E.} \bibnamefont{Kidder}},
  \bibinfo{author}{\bibfnamefont{L.}~\bibnamefont{Lindblom}},
  \bibinfo{author}{\bibfnamefont{G.}~\bibnamefont{Lovelace}}, \bibnamefont{and}
  \bibinfo{author}{\bibfnamefont{M.~A.} \bibnamefont{Scheel}},
  \bibinfo{journal}{Class.\ Quantum Grav.} \textbf{\bibinfo{volume}{24}},
  \bibinfo{pages}{S59} (\bibinfo{year}{2007}), \eprint{gr-qc/0702106}.

\bibitem[{\citenamefont{Tichy and Marronetti}(2010)}]{Tichy:2010qa}
\bibinfo{author}{\bibfnamefont{W.}~\bibnamefont{Tichy}} \bibnamefont{and}
  \bibinfo{author}{\bibfnamefont{P.}~\bibnamefont{Marronetti}}
  (\bibinfo{year}{2010}), \eprint{1010.2936}.

\bibitem[{\citenamefont{Buonanno et~al.}(2011)\citenamefont{Buonanno, Kidder,
  Mrou\'{e}, Pfeiffer, and Taracchini}}]{Buonanno:2010yk}
\bibinfo{author}{\bibfnamefont{A.}~\bibnamefont{Buonanno}},
  \bibinfo{author}{\bibfnamefont{L.~E.} \bibnamefont{Kidder}},
  \bibinfo{author}{\bibfnamefont{A.~H.} \bibnamefont{Mrou\'{e}}},
  \bibinfo{author}{\bibfnamefont{H.~P.} \bibnamefont{Pfeiffer}},
  \bibnamefont{and}
  \bibinfo{author}{\bibfnamefont{A.}~\bibnamefont{Taracchini}},
  \bibinfo{journal}{Phys.Rev.} \textbf{\bibinfo{volume}{D83}},
  \bibinfo{pages}{104034} (\bibinfo{year}{2011}), \eprint{1012.1549}.

\bibitem[{\citenamefont{Carminati and McLenaghan}(1991)}]{Carminati:1991}
\bibinfo{author}{\bibfnamefont{J.}~\bibnamefont{Carminati}} \bibnamefont{and}
  \bibinfo{author}{\bibfnamefont{R.}~\bibnamefont{McLenaghan}},
  \bibinfo{journal}{J. Math. Phys.} \textbf{\bibinfo{volume}{32}},
  \bibinfo{pages}{3135} (\bibinfo{year}{1991}).

\bibitem[{\citenamefont{Szekeres}(1965)}]{Szekeres1965}
\bibinfo{author}{\bibfnamefont{P.}~\bibnamefont{Szekeres}},
  \bibinfo{journal}{Journal of Mathematical Physics}
  \textbf{\bibinfo{volume}{6}}, \bibinfo{pages}{1387} (\bibinfo{year}{1965}),
  \urlprefix\url{http://link.aip.org/link/?JMP/6/1387/1}.

\bibitem[{\citenamefont{Baker and Campanelli}(2000)}]{Baker:2000zm}
\bibinfo{author}{\bibfnamefont{J.~G.} \bibnamefont{Baker}} \bibnamefont{and}
  \bibinfo{author}{\bibfnamefont{M.}~\bibnamefont{Campanelli}},
  \bibinfo{journal}{Phys.Rev.} \textbf{\bibinfo{volume}{D62}},
  \bibinfo{pages}{127501} (\bibinfo{year}{2000}), \eprint{gr-qc/0003031}.

\bibitem[{\citenamefont{Zhang et~al.}(2011)\citenamefont{Zhang, Brink,
  Szil\'{a}gyi, and Lovelace}}]{QuasiKinn}
\bibinfo{author}{\bibfnamefont{F.}~\bibnamefont{Zhang}},
  \bibinfo{author}{\bibfnamefont{J.}~\bibnamefont{Brink}},
  \bibinfo{author}{\bibfnamefont{B.}~\bibnamefont{Szil\'{a}gyi}},
  \bibnamefont{and} \bibinfo{author}{\bibfnamefont{G.}~\bibnamefont{Lovelace}}
  (\bibinfo{year}{2011}), \bibinfo{note}{in preparation}.

\bibitem[{\citenamefont{Nichols et~al.}(2011)\citenamefont{Nichols, Owen,
  Zhang, Zimmerman, Brink, Chen, Kaplan, Lovelace, Matthews, Scheel
  et~al.}}]{Nichols:2011pu}
\bibinfo{author}{\bibfnamefont{D.~A.} \bibnamefont{Nichols}},
  \bibinfo{author}{\bibfnamefont{R.}~\bibnamefont{Owen}},
  \bibinfo{author}{\bibfnamefont{F.}~\bibnamefont{Zhang}},
  \bibinfo{author}{\bibfnamefont{A.}~\bibnamefont{Zimmerman}},
  \bibinfo{author}{\bibfnamefont{J.}~\bibnamefont{Brink}},
  \bibinfo{author}{\bibfnamefont{Y.}~\bibnamefont{Chen}},
  \bibinfo{author}{\bibfnamefont{J.}~\bibnamefont{Kaplan}},
  \bibinfo{author}{\bibfnamefont{G.}~\bibnamefont{Lovelace}},
  \bibinfo{author}{\bibfnamefont{K.~D.} \bibnamefont{Matthews}},
  \bibinfo{author}{\bibfnamefont{M.~A.} \bibnamefont{Scheel}},
  \bibnamefont{et~al.} (\bibinfo{year}{2011}), \eprint{1108.5486}.

\bibitem[{\citenamefont{Lovelace et~al.}(2011)\citenamefont{Lovelace, Scheel,
  and Szilagyi}}]{Lovelace:2010ne}
\bibinfo{author}{\bibfnamefont{G.}~\bibnamefont{Lovelace}},
  \bibinfo{author}{\bibfnamefont{M.~A.} \bibnamefont{Scheel}},
  \bibnamefont{and} \bibinfo{author}{\bibfnamefont{B.}~\bibnamefont{Szilagyi}},
  \bibinfo{journal}{Phys.\ Rev.\ D} \textbf{\bibinfo{volume}{83}},
  \bibinfo{pages}{024010} (\bibinfo{year}{2011}), \eprint{1010.2777}.

\bibitem[{\citenamefont{Hesthaven}(1997)}]{Hesthaven1997}
\bibinfo{author}{\bibfnamefont{J.~S.} \bibnamefont{Hesthaven}},
  \bibinfo{journal}{SIAM J. Sci. Comput.} \textbf{\bibinfo{volume}{18}},
  \bibinfo{pages}{658} (\bibinfo{year}{1997}).

\bibitem[{\citenamefont{Hesthaven}(1999)}]{Hesthaven1999}
\bibinfo{author}{\bibfnamefont{J.~S.} \bibnamefont{Hesthaven}},
  \bibinfo{journal}{SIAM J. Sci. Comput.} \textbf{\bibinfo{volume}{20}},
  \bibinfo{pages}{62} (\bibinfo{year}{1999}).

\bibitem[{\citenamefont{Hesthaven}(2000)}]{Hesthaven2000}
\bibinfo{author}{\bibfnamefont{J.~S.} \bibnamefont{Hesthaven}},
  \bibinfo{journal}{Appl. Num. Math.} \textbf{\bibinfo{volume}{33}},
  \bibinfo{pages}{23} (\bibinfo{year}{2000}).

\bibitem[{\citenamefont{Gottlieb and Hesthaven}(2001)}]{Gottlieb2001}
\bibinfo{author}{\bibfnamefont{D.}~\bibnamefont{Gottlieb}} \bibnamefont{and}
  \bibinfo{author}{\bibfnamefont{J.~S.} \bibnamefont{Hesthaven}},
  \bibinfo{journal}{J. Comput. Appl. Math.} \textbf{\bibinfo{volume}{128}},
  \bibinfo{pages}{83} (\bibinfo{year}{2001}), ISSN \bibinfo{issn}{0377-0427},
  \urlprefix\url{http://dx.doi.org/10.1016/S0377-0427(00)00510-0}.

\bibitem[{\citenamefont{Lindblom
  et~al.}(2006{\natexlab{b}})\citenamefont{Lindblom, Scheel, Kidder, Owen, and
  Rinne}}]{Lindblom2006}
\bibinfo{author}{\bibfnamefont{L.}~\bibnamefont{Lindblom}},
  \bibinfo{author}{\bibfnamefont{M.~A.} \bibnamefont{Scheel}},
  \bibinfo{author}{\bibfnamefont{L.~E.} \bibnamefont{Kidder}},
  \bibinfo{author}{\bibfnamefont{R.}~\bibnamefont{Owen}}, \bibnamefont{and}
  \bibinfo{author}{\bibfnamefont{O.}~\bibnamefont{Rinne}},
  \bibinfo{journal}{Class.\ Quantum Grav.} \textbf{\bibinfo{volume}{23}},
  \bibinfo{pages}{S447} (\bibinfo{year}{2006}{\natexlab{b}}).

\bibitem[{\citenamefont{Scheel et~al.}(2006)\citenamefont{Scheel, Pfeiffer,
  Lindblom, Kidder, Rinne, and Teukolsky}}]{Scheel2006}
\bibinfo{author}{\bibfnamefont{M.~A.} \bibnamefont{Scheel}},
  \bibinfo{author}{\bibfnamefont{H.~P.} \bibnamefont{Pfeiffer}},
  \bibinfo{author}{\bibfnamefont{L.}~\bibnamefont{Lindblom}},
  \bibinfo{author}{\bibfnamefont{L.~E.} \bibnamefont{Kidder}},
  \bibinfo{author}{\bibfnamefont{O.}~\bibnamefont{Rinne}}, \bibnamefont{and}
  \bibinfo{author}{\bibfnamefont{S.~A.} \bibnamefont{Teukolsky}},
  \bibinfo{journal}{Phys.\ Rev.\ D} \textbf{\bibinfo{volume}{74}},
  \bibinfo{pages}{104006} (\bibinfo{year}{2006}).

\bibitem[{\citenamefont{Hemberger et~al.}(2013)\citenamefont{Hemberger, Scheel,
  Kidder, Szil\'agyi, Lovelace, Taylor, and Teukolsky}}]{Hemberger:2012jz}
\bibinfo{author}{\bibfnamefont{D.~A.} \bibnamefont{Hemberger}},
  \bibinfo{author}{\bibfnamefont{M.~A.} \bibnamefont{Scheel}},
  \bibinfo{author}{\bibfnamefont{L.~E.} \bibnamefont{Kidder}},
  \bibinfo{author}{\bibfnamefont{B.}~\bibnamefont{Szil\'agyi}},
  \bibinfo{author}{\bibfnamefont{G.}~\bibnamefont{Lovelace}},
  \bibinfo{author}{\bibfnamefont{N.~W.} \bibnamefont{Taylor}},
  \bibnamefont{and} \bibinfo{author}{\bibfnamefont{S.~A.}
  \bibnamefont{Teukolsky}}, \bibinfo{journal}{Class.\ Quantum Grav.}
  \textbf{\bibinfo{volume}{30}}, \bibinfo{pages}{115001}
  (\bibinfo{year}{2013}), \eprint{1211.6079},
  \urlprefix\url{http://stacks.iop.org/0264-9381/30/i=11/a=115001}.

\bibitem[{\citenamefont{Buchman and Sarbach}(2006)}]{Buchman2006}
\bibinfo{author}{\bibfnamefont{L.~T.} \bibnamefont{Buchman}} \bibnamefont{and}
  \bibinfo{author}{\bibfnamefont{O.~C.~A.} \bibnamefont{Sarbach}},
  \bibinfo{journal}{Class.\ Quantum Grav.} \textbf{\bibinfo{volume}{23}},
  \bibinfo{pages}{6709} (\bibinfo{year}{2006}),
  \urlprefix\url{http://stacks.iop.org/0264-9381/23/6709}.

\end{thebibliography}
\end{document}